\documentclass[
aps,
prx,
amsmath,amssymb,
superscriptaddress,
reprint,
nofootinbib,
floatfix,
longbibliography,
noeprint
]{revtex4-2}

\usepackage{accents}
\newcommand{\dbtilde}[1]{\accentset{\approx}{#1}}

\usepackage{mathtools}
\usepackage[english]{babel}
\usepackage[utf8]{inputenc}
\usepackage[T1]{fontenc}
\usepackage{physics}
\usepackage{amsmath}
\usepackage{mathrsfs}
\usepackage{bbm}
\usepackage{bm}
\usepackage{enumerate}
\usepackage{graphicx}
\usepackage[dvipsnames]{xcolor}
\usepackage{float}
\usepackage{subdepth}
\usepackage{makecell}
\usepackage{fnpct}
\usepackage{booktabs}
\usepackage{multirow}
\usepackage{setspace}
\usepackage[math]{cellspace}
\usepackage{subfigure}
\usepackage{mathtools}
\usepackage{soul}

\usepackage[colorlinks,
linkcolor=BrickRed,
citecolor=MidnightBlue,
urlcolor=MidnightBlue,
bookmarks=true,
bookmarksopen=true,
bookmarksnumbered=true,
]{hyperref}

\renewcommand{\hat}{\widehat}
\newcommand{\eref}[1]{(\ref{#1})}
\newcommand{\dob}{\begin{tabular}{c}}
\newcommand{\edob}{\end{tabular} }

\renewcommand{\Re}{\rm Re}
\newcommand{\nn}{\nonumber}

\renewcommand{\vec}[1]{\underline{#1}}
\renewcommand{\i}{\mathrm{i}}
\renewcommand{\(}{\left(}
\renewcommand{\)}{\right)}
\newcommand{\id}{\mathbbm{1}}

\newcommand{\be}{\begin{eqnarray}}
\newcommand{\bea}{\begin{eqnarray}}
\newcommand{\eea}{\end{eqnarray}}
\newcommand{\beq}{\begin{equation}}
\newcommand{\ee}{\end{eqnarray}}
\newcommand{\eeq}{\end{equation}}

 \newcounter{ls}

\newcounter{amg}

\newcounter{jvc}

\newcounter{zl}

\def\equationautorefname~#1\null{Eq. (#1)\null}

\newcommand{\appref}[1]{\hyperref[#1]{App.~\ref*{#1}}}
\usepackage[all]{hypcap} %
\usepackage{physics}

\graphicspath{ {./figures/} }

\begin{document}
	
\title{Lindblad many-body scars}
\author{Antonio M. Garc\'\i a-Garc\'\i a}
\email{amgg@sjtu.edu.cn}
\affiliation{Shanghai Center for Complex Physics,
	School of Physics and Astronomy, Shanghai Jiao Tong
	University, Shanghai 200240, China}
\author{Zhongling Lu}
\email{zhongling_lu@sjtu.edu.cn}
\affiliation{Shanghai Center for Complex Physics,
	School of Physics and Astronomy, Shanghai Jiao Tong
	University, Shanghai 200240, China}
\affiliation{Zhiyuan College, Shanghai Jiao Tong University, Shanghai 200240, China}
\author{Lucas S\'a}
\email{ld710@cam.ac.uk}
\affiliation{TCM Group, Cavendish Laboratory, University of Cambridge, JJ Thomson Avenue, Cambridge CB3 0HE, UK\looseness=-1}

\author{Jacobus J. M. Verbaarschot}
\email{jacobus.verbaarschot@stonybrook.edu}

\affiliation{Center for Nuclear Theory and Department of Physics and Astronomy, Stony Brook University, Stony Brook, New York 11794, USA}

\date{\today}

\begin{abstract}
Quantum many-body scars have received much recent attention for being both intriguing non-ergodic states in otherwise quantum chaotic systems and promising candidates to encode quantum information efficiently. So far, these studies have mostly been restricted to Hermitian systems. 
Here, we study many-body scars in many-body quantum chaotic systems coupled to a Markovian bath, which we term Lindblad many-body scars. They are defined as
simultaneous eigenvectors of the Hamiltonian and dissipative parts of the vectorized Liouvillian. Importantly, because their eigenvalues are purely real, they are not related to revivals.
The number and nature of the scars depend on both the symmetry of the Hamiltonian and the choice of jump operators. For a dissipative four-body Sachdev-Ye-Kitaev (SYK) model with $N$ fermions, either Majorana or complex, we construct analytically some of these Lindblad scars while others could only be obtained numerically.
As an example of the former, we identify $N/2+1$ scars for complex fermions due to the $U(1)$ symmetry of the model and two scars for Majorana fermions as a consequence of the parity symmetry.
Similar results are obtained for a dissipative XXZ spin chain. 
We also characterize the physical properties of Lindblad scars. First, the operator size is independent of the disorder realization and has a vanishing variance. 
By contrast, the operator size for non-scarred states, believed to be quantum chaotic, is well described by a distribution centered around a specific size and a finite variance, which could be relevant for a precise definition of the eigenstate thermalization hypothesis in dissipative quantum chaos.
Moreover, the entanglement entropy of these scars has distinct features such as a strong dependence on the partition choice and, in certain cases, a large entanglement. 
\end{abstract}

\maketitle

\section{Introduction}

Quantum scars \cite{heller1984}, originally discovered 40 years ago, are eigenstates of quantum chaotic systems whose wave functions are strongly localized along an unstable periodic orbit. The initial concept of quantum scars emerged from the studies of classical-quantum correspondence in single-particle systems, such as quantum billiards, which exhibit distinct semiclassical limits and have been directly observed \cite{ge2024direct} in experiments. More recently, the study of scars has received an important boost in the context of many-body quantum systems. In this case, due to the absence of a clear semiclassical characterization (see, however, Refs.~\cite{evrard2024PRL,evrard2024PRB,pizzi2024,graf2024,pizzi2024trails,lu2025anti,hummel2023genuine} for recent developments along this line), quantum many-body scars
\cite{serbyn2021,bernien2017,bluvstein2021,turner2018,Pakrouski:2020hym} were identified, both theoretically and experimentally, by non-thermal oscillatory dynamics for some specific initial conditions related to the existence of towers of states with an equally spaced spectrum. 

More generally, many-body scar states can be defined as analytical eigenstates of a
system. 
Typically, one observes the coexistence of eigenstates that
are given by an explicit analytical expression, and eigenstates that can only be
obtained numerically which show typical characteristics of chaotic states such as avoided level crossings and eigenstate thermalization.
The coexistence of these two types of states was first identified
in three- and four-anyon systems in a harmonic well \cite{Sporre:1991qm,Sporre:1991pm,Sporre:1991ui}, while the discovery of integrable many-anyon states goes back much
further \cite{wu1984multiparticle}. Related questions were discussed in the context
of ``partial algebraization'' of the spectrum \cite{Shifman:1998si,Shifman:1998tj}.

In a closely related development, it was found \cite{shiraishi2017,moudgalya2018} that it was possible to engineer quantum chaotic Hamiltonians so that specific eigenstates
violate the eigenstate thermalization hypothesis (ETH) and therefore could also be considered as quantum scars.
Indeed, one of the main
theoretical motivations to study many-body scars was concerning the conditions for the violation of the ETH and more generally with the process of thermalization in closed quantum many-body systems.

Another major interest in quantum scars comes from quantum information applications \cite{chen2022error,cao2024}. Thermalization, which is generic in many-body quantum chaotic systems is fast which makes it difficult to use them for encoding and manipulating quantum information. Many-body localized systems do not suffer from this problem, but their low entanglement is a serious drawback for quantum information applications that typically require ample entanglement resources. 
In principle, it is plausible to expect that quantum scars are also configurations with low entanglement. However,
it has been shown that this is not always the case. The so-called {\it rainbow scars} \cite{ramirez2015,langlett2022,ramirez2015,dong2023} can have an entanglement entropy (EE) that satisfies a volume law \cite{langlett2022}. More interestingly, the level of entanglement can be modified \cite{dong2023}, at least to some extent, by tuning the parameters of the model; see Ref.~\cite{bernevig2022} for a recent review.  

So far the discussion has been restricted to Hermitian systems. However, in the aforementioned quantum information setting, the unavoidable measuring protocol makes the dynamics effectively non-Hermitian. The same applies if we consider the effect of an environment which is described by a similar formalism \cite{breuer2002}. Therefore, a natural and rather fundamental problem to investigate is the existence and characteristics of quantum scars in this more general setting.

The literature \cite{buvca2019NatComm,chen2023,shen2024,leeuw2024,cao2024,hamazaki2024,abanin2025} on this problem is still rather scarce. 
The skin-effect \cite{shen2024} has been employed to identify a scarred eigenstate in a non-Hermitian projected spin model. The same model was used \cite{chen2023} to characterize non-Hermitian scars as states with low EE and leading to revivals. Very recently, Refs.~\cite{wang2024,jiang2025} reported the existence of revivals due to many-body scars in a Lindblad setting. Moreover, the steady state of a vectorized Lindbladian at finite temperature, a thermofield double (TFD) state, has been characterized as a many-body scar \cite{teretenkov2024}. However, even basic questions like the characterization of many-body scars in dissipative quantum systems and their entanglement features are still poorly understood. Below, we generalize the definition of the TFD scar of Ref.~\cite{teretenkov2024} to more general states anywhere in the spectrum and provide the first extensive study of the properties of these states. 
 
The main goal of this paper is to address the above questions by providing a detailed description of scars in vectorized Liouvillians that describe the dynamics of many-body quantum chaotic systems coupled to a Markovian bath. This would open new research avenues in dissipative quantum chaos where even basic questions such as its very definition \cite{garcia2023a} or the precise role and meaning of the ETH \cite{cipolloni2024,mao2024,roy2023,hamazaki2022arXiv,almeida2025} are still being debated.
In the context of quantum information, it would allow for a more precise assessment of the potential of quantum scars as entanglement resources once the unavoidable process of measurement is considered. 

For these purposes, we shall employ the so-called Lindbladian Sachdev-Ye-Kitaev (SYK) model \cite{sa2022,kulkarni2022,garcia2022e,kawabata2022,kawabata2023symmetry,garcia2023c} that describes the dynamics of an SYK model \cite{bohigas1971,french1970,french1971,gervois:1972owy,sachdev1993,benet2003,kitaev2015,maldacena2016} in contact with a Markovian bath. The SYK model is zero-dimensional, which makes it possible to reach larger sizes while keeping \cite{maldacena2015,maldacena2016,garcia2016} all distinct features of many-body quantum chaos. Moreover,
it is a toy model for holography \cite{kitaev2015,maldacena2016}; in a certain region of parameters, our results may thus have a gravity dual interpretation. We shall see explicitly that the main results of the paper are not specific to the SYK model but also apply to other strongly interacting systems such as spin models.       
We start with the definition and general discussion of scar states.             

\section{Vectorized Lindbladian and Many-body scars} \label{sec:operator scars}
We explore the existence and characterization of many-body scars in quantum chaotic systems coupled to a Markovian bath. The time evolution of the density matrix of the system $\hat{\rho}$ is governed by a Liouvillian ${\cal L}$ of Lindblad form~\cite{lindblad1976,gorini1976}, 
\begin{equation} \label{eq:lindblad_operator_form}
  \partial_t \hat{\rho}= \mathcal{L}(\hat{\rho})= -\i[\hat{H}, \hat{\rho}]+ \mu \sum_\alpha
  \left(\hat{L}_\alpha \hat{\rho} \hat{L}_\alpha^{\dagger}-\frac{1}{2}\left\{\hat{L}_\alpha^{\dagger} \hat{L}_\alpha, \hat{\rho}\right\}\right),
\end{equation}
where $\hat{H}$ is the many-body quantum chaotic Hamiltonian and ${\hat L_\alpha}$ are the jump operators that characterize coupling to the bath with strength $\mu$.
To study the evolution of the density matrix, we follow the standard procedure \cite{sieberer2016keldysh} of writing the rows of the matrix as successive columns.
This is known as the operator-state mapping, i.e. an operator $\hat{\mathcal{O}}$ in the original Hilbert space $\mathcal{H}$ is mapped to a state $\ket{\mathcal{O}}$ in a doubled Hilbert space $\mathcal{H}_2=\mathcal{H} \otimes \mathcal{H}^* \equiv \mathcal{H}^L \otimes \mathcal{H}^R$, where $^*$ stands for the dual space. 
The resulting Liouvillian, usually referred to as the vectorized Liouvillian, is given by   
   
\begin{equation}\label{eq:vectorized_Liouvillian}
	\mathcal{L}=-\i H_0 + H_I,
\end{equation} 
where $H_0 = H^L - H^R$ and $H_I$ represents the dissipative term. 
Here, $H^L$ and $H^R$ can also be interpreted as  the Hamiltonian on the backward and forward Keldysh contours, respectively.
For specific choices of $\hat{H}$, such as a SYK model, this general construction may be dual~\cite{turiaci2019} to a weakly perturbed global de Sitter geometry in two dimensions, which might be unstable toward the formation of Keldysh wormholes~\cite{garcia2021c}. 
The precise form of $H^L$, $H^R$, and $H_I$ depends on the choice of the mapping (i.e., vectorization). We discuss different possibilities below, which we show to correspond to distinct basis choices for the doubled Hilbert space.

A Hamiltonian analog of Eq.~(\ref{eq:vectorized_Liouvillian}) with $\i H^L \to H^L$, $-\i H^R \to H^R$, and $H_I$ Hermitian, was introduced \cite{langlett2022,dong2023} to construct so-called {\it rainbow scars} \cite{ramirez2015}. Similarly,
many-body scars can also be constructed for the Liouvillian $\mathcal{L}$ Eq.~(\ref{eq:vectorized_Liouvillian}) as follows.
The TFD at infinite temperature
 \cite{maldacena2003},
 	\begin{equation}
 	|0\rangle=\frac{1}{D} \sum_{i=1}^{D}| s_i^L\rangle | s_i^R\rangle,
 \end{equation}
where $|s_i^L\rangle$ ($|s_i^R\rangle$) are the eigenstates of $H^L$ ($H^R$) and $D$ is the Hilbert space dimension, is an eigenstate\footnote{The term `eigenstate' is used throughout this work to refer to the eigenmodes of the vectorized Liouvillian. They are equivalently eigenoperators of the Lindblad equation in the original Hilbert space due to the operator-state mapping.} of 
$H_0 = H^L - H^R$ with eigenvalue zero due to the relative minus sign between $H^L$ and $H^R$. 
Contrary to the Hermitian setting of Ref.~\cite{langlett2022}, where one needs to introduce the relative sign by hand, here it is inherent to the Lindblad formalism that describes a single system coupled to a bath.  

If the TFD is also an eigenstate of the interacting part $H_I$, then it becomes an eigenstate of the whole Liouvillian. The TFD corresponds to the vectorization of the identity matrix and hence it is an eigenstate of $H_I$ with real eigenvalue $\mu \lambda$ if $\sum_\alpha [\hat{L}_\alpha, \hat{L}_\alpha^\dagger]=\lambda \id$. If $\lambda=0$, the TFD corresponds to the steady state of the Liouvillian, which always exists in the spectrum.

The above example shows a trivial example of a scar~\cite{teretenkov2024}. Inspired by it, we define a \emph{Lindblad} many-body scar as a simultaneous eigenstate of $H_0$ and $H_I$ in Eq.~(\ref{eq:vectorized_Liouvillian}) whose eigenvalue is the same for an entire family of Liouvillians~\cite{teretenkov2024}. For example, if the Liouvillian is disordered, the latter restriction guarantees that the eigenvalue of the scar state is independent of the disorder realization. However, in general, the scarred eigenvector will depend on the disorder realization.
For concreteness, we will focus only on the case where the scar vanishes under the action of $H_0$ and hence its eigenvalue is completely determined by the dissipative part $H_I$.
The same definition can be formulated in the original operator language of the Lindblad equation~\cite{teretenkov2024}. Let $\hat{\mathcal{O}}$ be the operator that corresponds to the vectorized state $\ket{\mathcal{O}}$. $\ket{\mathcal{O}}$ is a scar if $\hat{\mathcal{O}}$ satisfies the following two conditions: 
\begin{enumerate}[(i)]
	\item $[\hat{H},\hat{\mathcal{O}}]=0,$
	\item $\sum_\alpha \left(\hat{L}_\alpha \hat{\mathcal{O}} \hat{L}_\alpha^{\dagger}
	-\frac{1}{2}\left\{\hat{L}_\alpha^{\dagger} \hat{L}_\alpha, \hat{\mathcal{O}}\right\}\right) =\eta \hat{\mathcal{O}},$
\end{enumerate}
where $\eta$ is a constant. We thus have that $\mathcal{L}[\hat{\mathcal{O}}] = \mu\eta \hat{\mathcal{O}}$.
In the remainder of the paper, we will restrict our attention to Hermitian jump operators, $\hat{L}_\alpha=\hat{L}^\dagger_\alpha$ that satisfy $\hat{L}_\alpha^2=a \id$ with constant $a$, in which case $H_I$ is Hermitian and the TFD becomes the steady state. In this case, condition (ii) simplifies to 
\begin{enumerate}[(ii')]
	\item $\sum_\alpha \hat{L}_\alpha\mathcal{O}\hat{L}_\alpha = \eta' \hat{\mathcal{O}},$
\end{enumerate} 
for some real constant $\eta'$ related to $\eta$ by a shift.

The main feature of the scar states is that their eigenvalues can be obtained analytically
and do not depend on
the details of $\hat{H}$. To see this,
first note that for \textit{any} matrix, the real and imaginary part of the eigenvalues are related to expectation values of the Hermitian and anti-Hermitian parts of the matrix. Indeed, for an eigenstate $\ket{k}$, let $\mathcal{L} \ket{k} = \lambda_k\ket{k}$, where
$\mathcal{L} = -\i H_0 + H_I$
is decomposed into its Hermitian part $H_I$ and anti-Hermitian
part $-\i H_0$ and $\lambda_k=a_k+\i b_k$ with $a_k$, $b_k \in \mathbb{R}$.
Then, this leads to $-\i \bra{k} H_0 \ket{k} + \bra{k} H_I \ket{k} = a_k + \i b_k$, and because $a_k$ and $b_k$ are real and $H_0$ and $H_I$ are Hermitian, it follows that $a_k = \bra{k}H_I\ket{k}$, $b_k = - \bra{k}H_0\ket{k}$. 
This is not helpful for generic eigenstates of the Liouvillian. However, scar states are also eigenstates of $H_0$ and $H_I$ individually, and therefore, the eigenvalues of scar states can be obtained without diagonalization. Because of condition (i), we have $b_k =0$, and by condition (ii'), $a_k=\mu \eta$ for any choices of $H_0$ consistent with condition~(i). In Sec.~\ref{sec:operator_size}, we will see that this construction relates to the operator size in open quantum systems.

Our definition of Lindblad scars is related to that of coherent many-body eigenstates considered in Ref.~\cite{buca2012}. Indeed, we can relax condition (i) to $[\widehat{H},\widehat{{\mathcal{O}}}]=\omega \widehat{{\mathcal{O}}}$ with real $\omega$, allowing for oscillatory scar states. If we then set $\eta=0$, we obtain non-steady eigenstates that survive in the long-time limit (the system does not relax to a steady state). Such eigenstates are similar, but not identical to dark states~\cite{diehl2008,kraus2008} and decoherence-free subspaces~\cite{lidar1998,lidar2003}, since they typically correspond to isolated eigenvalues and do not span a subspace of the underlying Hilbert space of the isolated system. Moreover, we note that the phenomenology of scars with $\eta\neq0$ and $\eta=0$ is markedly different because of the decay induced by the finite real part of the eigenvalues when $\eta\neq0$. In the remainder of the paper we will consider only $\omega=0$, as required by condition (i); we emphasize that these Lindblad scars are in general not related to revivals, oscillatory dynamics, or decoherence-free subspaces.

As an illustration of the above analysis, we now identify many-body scars in a SYK model coupled to a Markovian bath for both Majorana and complex fermions.

\section{Lindblad scars in the SYK model}

\subsection{Dissipative Majorana SYK model}

We first investigate the Lindbladian Majorana SYK model \cite{sa2022,kulkarni2022,garcia2022e,kawabata2022,kawabata2023symmetry,garcia2023c} described by the Liouvillian Eq.~(\ref{eq:lindblad_operator_form}) with 
\begin{equation}
	\label{eq:Maj_SYK_single_Hamil}
	\hat{H}=-i^{\frac{q}{2}} \sum_{1 \leq i_1<i_2 \cdots<i_q \leq N} K_{i_1 i_2 \cdots i_q} \psi_{i_1} \psi_{i_2} \cdots \psi_{i_q}
\end{equation}
where $\psi_i$ are Majorana fermions, $\left\{\psi_i, \psi_j\right\}=\delta_{i j}$, $K_{i_1 i_2 \cdots i_q}$ are Gaussian random couplings with zero mean and variance
$2^{q-1} (q-1)!N^{1-q}/q$, $N$ is even and $q$ is an even integer.
The Markovian bath enters in the Lindblad equation through 
the jump operators, which we choose as $\hat{L}_i= \psi_i$.
Because $\hat{L}_i^\dagger \hat{L}_i=\id/2 $, the term
$\{\sum_\alpha{\hat{L}_\alpha^\dagger \hat{L}_\alpha,\hat{\rho}}\}$ is simply an overall shift of the Lindbladian.

The vectorization of the Liouvillian is not unique. While different vectorization schemes correspond to unitary transformations of the Liouvillian and therefore leave, e.g., the spectrum invariant, they give rise to different partitions of the many-body Hilbert space and, therefore, affect quantities that are sensitive to them such as the size discussed in Sec.~\ref{sec:size} and 
the EE discussed in Sec.~\ref{sec:EE}. We start this section by precisely discussing different choices of vectorization.

A commonly used~\cite{garcia2022e} set of left and right Majorana fermions is given by
 \begin{equation}
 	\label{eq:fermions1}
 \chi_k^L = \psi_k\otimes \id, \qquad  \chi_k^R =  \hat{P} \otimes  \psi_k, 
 \end{equation}
where the parity operator $\hat{P} = 2^{N/2} i^{N(N-1)/2}\prod_{k=1}^N \psi_k$ anti-commutes with all $\psi_k$ (i.e., $\hat{P}$ is the $\Gamma_5$ matrix for the Euclidean $N$-dimensional gamma matrices that represent the fermions).
With this representation, the left, right, and interaction terms in the Liouvillian read~\cite{garcia2022e}:
\begin{align} \label{eq:HI}
	&H^L=-\i^{\frac{q}{2}} \sum_{1 \leq i_1<\cdots<i_q \leq N} K_{i_1 i_2 \cdots i_q} \chi^L_{i_1} \cdots \chi^L_{i_q},
	\\
	&H^R=-\i^{\frac{q}{2}}(-1)^{q/2}\sum_{1 \leq i_1<\cdots<i_q \leq N} K_{i_1 i_2 \cdots i_q} \chi^R_{i_1} \cdots \chi^R_{i_q},
	\\
	&H_I=\i\mu\sum_i \chi_i^L\chi_i^R-\frac{N\mu}{2}.
        \label{hi}
\end{align}
We can construct a different set of Majoranas by the unitary transformation
 \be
 U =e^{\i\pi \hat{P} /4} \otimes \hat{C},
 \ee
 with $\hat{C}$ the unitary charge conjugation matrix satisfying $\hat{C}\psi_k \hat{C}^{-1} =\psi_k^*$. 
 This gives the Majorana fermions
 \be
 \widetilde \chi_k^L &=& U \chi_k^L U^{-1} = \i \hat{P} \psi_k\otimes \id,\\
 \widetilde \chi_k^R &=& U \chi_k^R U^{-1} = \hat{P} \otimes \psi_k^*.
 \ee
 The left-right coupling can be rewritten as
 \begin{equation}
 \begin{split}
 U\i\chi^L_k \chi^R_k U^{-1} 
 &= -(\hat{P}\psi_k\otimes \id) (\hat{P}\otimes\psi_k^*) 
 \\
 &=(\psi_k \otimes \id) (\id\otimes \psi_k^*) \equiv\dbtilde{\chi}_k^L \dbtilde{\chi}_k^R,
 \end{split}
 \end{equation}
 where we have introduced the ``pseudo-fermions''
 \be
 \label{eq:vectorization_Maj}
\dbtilde \chi^L_k  = \psi_k \otimes \id, \qquad
\dbtilde\chi^R_k =\id  \otimes \psi_k^*.
\ee
These left and right fermions anti-commute among themselves, but all left fermions commute with all right fermions. 
However, this not problematic because the Liouvillians in the two representations are related by a unitary transformation.
The interaction $H_I$ thus reads
\begin{equation}\label{eq:interact_term}
	H_I= \mu \sum_{k=1}^N \dbtilde{\chi}_{k}^{L}  \dbtilde{\chi}_{k}^{R} -\frac{N\mu}{2}.
\end{equation}
It is clear that $H^L$ and $H^R$ can also be rewritten in terms of these pseudo-fermions:
\begin{align}
\sum_{i_1\leq\cdots \leq i_q}  \chi^L_{i_1}\cdots\chi^L_{i_q}
&=&
\sum_{i_1\leq\cdots \leq i_q}  \dbtilde\chi^L_{i_1}\cdots\dbtilde\chi^L_{i_q} 
\\
(-1)^{\frac q2}\sum_{i_1\leq\cdots \leq i_q}  \chi^R_{i_1}\cdots\chi^R_{i_q}
&=&
\sum_{i_1\leq\cdots \leq i_q}  \dbtilde\chi^R_{i_1}\cdots\dbtilde\chi^R_{i_q} .
\end{align}
All vectorization schemes give the same results for quantities that are basis invariant, such as the spectrum. However, as we will see in Sec.~\ref{sec:EE}, it is advantageous to use these pseudo-fermions for the calculation of the EE.

The SYK Hamiltonian commutes only with the parity operator (and trivially with itself) and hence $\hat{\mathcal{O}}=\hat{H}$, $\hat{P}$, and $\hat{H}\hat{P}$ satisfy the scar condition (i). Moreover, they also satisfy condition (ii'):
\begin{align}
\label{eq:comm_P}
\sum_\alpha \psi_\alpha \hat{P} \psi_\alpha &= - \frac{N}{2} \hat{P}, \\
\label{eq:comm_H}
\sum_\alpha \psi_\alpha \hat{H} \psi_\alpha &= \left(\frac{N}{2}-q\right) \hat{H},\\
\label{eq:comm_HP}
\sum_\alpha \psi_\alpha \hat{H}\hat{P} \psi_\alpha &= -\left(\frac{N}{2}-q\right) \hat{H}\hat{P}.
\end{align}
This allows the construction of three more scar states in addition to the identity
operator: $\hat{P}$, $\hat{H}$, and $\hat{P}\hat{H}$, with eigenvalues $-N\mu$, $-q\mu$, and $-(N-q)\mu$, respectively.

The same calculation can be done in the vectorized model. For instance, the parity operator is mapped to the left parity, $P^L = 2^{N/2} \i^{N(N-1)/2}\prod_{k=1}^N \chi_k^L$, acting on the TFD $|0\rangle $. Then,
\begin{align}
&\i\sum_k \chi^L_k \chi^R_k P^L |0\rangle
=\i\sum_k \chi^L_k P^L \chi^R_k |0\rangle
\nn \\
=&\sum_k \chi^L_k P^L \chi^L_k |0\rangle = -\frac N2 P^L |0\rangle,
\end{align}
where in the second step we have used operator reflection~\cite{garcia2022e}. Considering the constant term of the vectorized operator, we find a scar state with
eigenvalue $-\mu N$.
The second scar, $\hat{\mathcal{O}}=\hat{H}$,
can be implemented in a similar fashion for the vectorized operator:
\begin{align}
&\i\sum_k \chi^L_k \chi^R_k H^L |0\rangle
= \i\sum_k \chi^L_k H^L \chi^R_k |0\rangle
\nn\\
=&\sum_k \chi^L_k H^L \chi^L_k |0\rangle = \left(\frac N2 -q\right) H^L |0\rangle.
\end{align}
Considering the constant term in the vectorized Liouvillian, this gives a
scar state with eigenvalue $-q\mu$. Similarly, the last symmetry gives an
eigenstate with eigenvalue $-(N-q)\mu$.
We thus have found four scar states:
\be
|0\rangle, \quad P^L |0\rangle, \quad H^L |0\rangle,  \quad H^L P^L |0\rangle.
\ee
Note that $(H^L)^2$ and higher powers of $H^L$ do not satisfy the symmetry relations (ii'). Therefore, there are no additional scars in the dissipative Majorana SYK model.

\begin{figure*}[t]
	\includegraphics[width=0.32\textwidth]{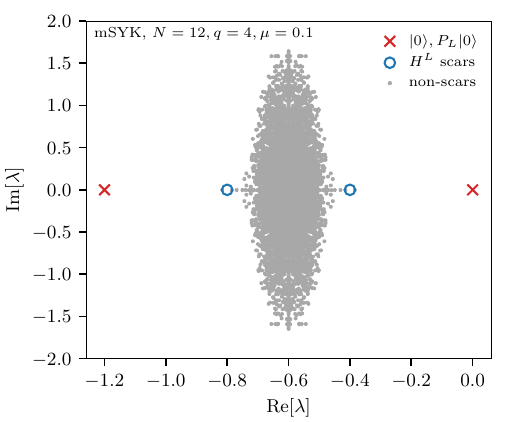}
	\includegraphics[width=0.32\textwidth]{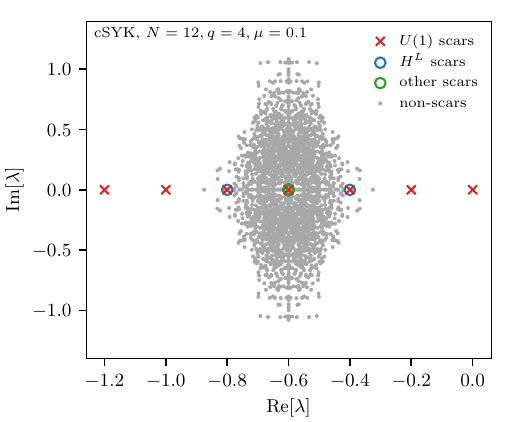}
        \includegraphics[width=0.32\textwidth] {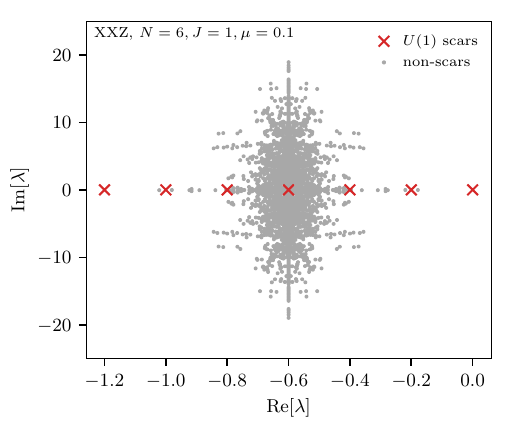}
	\caption{Spectrum $\lambda$ of one realization of the SYK Liouvillian with linear jump operators, with $q=4$ and $N=12$,
(left and middle) and of the XXZ spin model  (right). 
	Non-scar states are denoted by gray dots.
	Left: Majorana SYK Hamiltonian Eq.~(\ref{eq:Maj_SYK_single_Hamil}), which has two parity (red crosses) and two $H^L$ scars (blue circles). Middle: complex SYK Hamiltonian Eq.~(\ref{eq:sykhc}), which displays $N/2+1$ $U(1)$ scars (red crosses) and two $H^L$ scars (blue circles), which are fully characterized in the text, as well as 14 degenerate ``other scars'' (green circles) for which we could not find a full characterization. Right: the XXZ model  \eref{eq:xxz_hamiltonian} with weak dissipation (note the scale of the $y$-axis).
    In this case we only found the $U(1)$ scars (red crosses).
    All scar eigenvalues have zero imaginary part and an analytically
    determined real part  that does not depend on the disorder realization of the Hamiltonian.   }
	\label{fig:complex_N=12_eigval_scatter}
\end{figure*}

To numerically confirm the analytical results, we computed the spectrum of the dissipative Majorana SYK model with linear jump operators for $N=12$ and $\mu=0.1$, see Fig.~\ref{fig:complex_N=12_eigval_scatter} (left).
The eigenvalues of the scar states are denoted by red crosses and blue circles  and occur at the   locations explained before.

\subsection{Dissipative complex SYK model}
\label{sec:complex_SYK}

To find a richer structure of scar states, we consider the complex SYK Hamiltonian with $U(1)$ symmetry. 
To this end, we introduce the projection operator into  a sector of fixed charge (i.e., particle number) $n$, 
\be
\hat{\mathbb{P}}_n=\hat{\mathbb{P}}_n^2=\hat{\mathbb{P}}_n^{\dagger}.
\ee
Explicitly, let us consider complex fermions $a_i=(\psi_{2i-1}-\i\psi_{2i})/\sqrt{2}$, whose number operator
\begin{equation}
	\hat{\mathcal{N}}
	= \sum_{i=1}^{N / 2} a^\dagger_i a_i 
	= \sum_{i=1}^{N / 2}\left(\frac12- \i\psi_{2 i-1} \psi_{2 i}\right)
\end{equation} 
has eigenvalues $n=0, \ldots, N / 2$.
The projection operator onto
the sector with charge $n$ is given by
\begin{equation}
\hat{\mathbb{P}}_n=\frac{1}{N / 2+1} \sum_{s=0}^{N / 2} \exp \left\{\frac{2 \pi\i s}{N / 2+1}(\hat{\mathcal{N}}-n)\right\}
\end{equation}
and the complex SYK model is then obtained by summing over the projection onto all sectors,
\begin{equation}\label{eq:sykhc}
\hat{H}_c=\sum_{n=0}^{N / 2} \hat{\mathbb{P}}_n \hat{H} \hat{\mathbb{P}}_n.
\end{equation}
In a representation where the number operator is diagonal, we conjecture that this Hamiltonian agrees with the antisymmetrized complex SYK Hamiltonian introduced in Ref.~\cite{Gu:2019jub}. This can be shown analytically for $q=2$, while for
$q=4$, we only have numerical evidence.
The jump operators are chosen the same as in the Majorana case, namely, $\hat{L}_i=\psi_i$. It is crucial that the jump operators do not respect the U(1) symmetry. If they did, the whole Liouvillian would have a strong \cite{buca2012} $U(1)$ symmetry and the symmetry operator would lead to block diagonalization of $\mathcal{L}$ instead of inducing a scar.

If the number operator of the $k$-th complex fermion is given by
\begin{equation}
	\label{eq:def_nk}
\hat{n}_k = a^\dagger_k a_k=\frac{1}{2}-\i\psi_{2k-1}\psi_{2k},
\end{equation}
then we define the $p$-tuple operator as
\begin{equation}
	\label{eq:tpl}
\hat{\mathcal{N}}_p = \sum_{k_1\ne k_2\ne\cdots \ne k_p} \({\hat n}_{k_1}-\frac 12\) \({\hat n}_{k_2} -\frac 12\) \cdots \({\hat n}_{k_p}-\frac 12\).
\end{equation}
These operators commute with $\hat{H}_c$ [condition (i)] and are acted on by the jump operators as [condition (ii')]
\begin{equation}
	\sum_k \psi_k \hat{\mathcal{N}}_p \psi_k = \left ( \frac N2  -2p \right) \hat{\mathcal{N}}_p,
	\label{sym}
\end{equation}	
where $p$ takes the values, $0,1,\dots,N/2$, resulting in $N/2+1$ scar states with eigenvalue $-2p\mu$.
For the vectorized Liouvillian, they are given by
\begin{equation}
	\mathcal{L}\mathcal{N}_p^L|0\rangle
	=-2p\mu\mathcal{N}_p^L|0\rangle.
\end{equation}
Then the U(1) scar states of the vectorized Liouvillian are given by $\mathcal{N}_p^L|0\rangle$.
We note that the $p$-tuple operators can be expressed as linear combinations of powers of the number operator,
\be
\hat{\mathcal{N}}_p = \sum_{k=0}^p  c_{pk} \hat{\mathcal{N}}^k,
\ee
with $c_{pk}$ combinatorial factors that can be calculated recursively.
Indeed, the operators $\hat{\mathcal{N}}_p$ defined in Eq.~$\eref{eq:tpl}$ satisfy the relation
\begin{align} \label{eq:cSYK_scars}
\hat{\mathcal{N}}_{p+1}=\(\hat{\mathcal{N}} -\frac{N}2 \) \hat{\mathcal{N}}_p -\frac p4\(N-p+1\) \hat{\mathcal{N}}_{p-1},
\end{align}
with initial conditions $ \hat{\mathcal{N}}_0 = \mathbbm{1}$ and $\hat{\mathcal{N}}_1 = \hat{\mathcal{N}}
-N/2$.
For $\mathcal{N}_2$ and $\mathcal{N}_3$ we obtain
\begin{align}
\hat{\mathcal{N}}_2&=\(\hat{\mathcal{N}} - \frac N2\)^2 - \frac N4,
\\ 
\hat{\mathcal{N}}_3&= \(\hat{\mathcal{N}} - \frac N2\)^3 - \(\frac 34 N -\frac12\)  \(\hat{\mathcal{N}} - \frac N2 \).
\end{align}

As was the case for Majorana fermions, there is an additional scar state at both $-q\mu$ and $-(N-q)\mu$, which are given by
\be
H_{c}^L |0\rangle, \qquad H_{c}^L P^L |0\rangle,
\ee
respectively.
The existence of these scars still follows from Eqs.~(\ref{eq:comm_H}) and (\ref{eq:comm_HP}). Indeed, the projectors $\hat{\mathbb{P}}_n$ discard non-charge conserving terms from $H$ but do not change the number $q$ of Majorana operators in each term in $\hat{H}_c$.
Therefore, the commutation relation with $H_I$ is left unchanged and these scars also exist in the U(1) conserving system.

Are these the only scar eigenstates of the Lindbladian complex SYK model?
We performed numerical checks for $N=12$, and selected the scar states by the requirement that they are joint eigenstates of
$H^L - H^R$ and $H_I$.
To obtain disorder-independent eigenvalues, they  must be eigenstates of $H^L - H^R$ with eigenvalue zero, so a necessary requirement is that the eigenvalues of $\mathcal{L}$ are equal to  $\{-N\mu, (-N+2)\mu,  \dots, 0\} $. 
When these states are non-degenerate,
they are also eigenstates of $H_I$ and are scar states with the same eigenvalue.
When they are degenerate, a complication arises since they can be linear combinations
of scar and non-scar states. To disentangle the scar and non-scar states, we determine the singular values
 of $\langle \alpha |H_I -2k\mu| 2k\rangle$,
with $ | 2k\rangle$ the eigenstates of $\mathcal{L}$ with eigenvalue $2k\mu$ and $|\alpha\rangle$ a complete set of states.
Only the states corresponding
to singular value zero are scar states.
Following this procedure, we find a single scar eigenstate for all $-2p\mu$ except for $-q\mu$, $-(N- q)\mu$, and $-N \mu/2$. 
At $- q\mu$, there are two scar eigenstates given by $\mathcal{N}_1^L|0\rangle$ and $\hat{H}_c^L|0\rangle$ as discussed above. The same applies for the two scars at $-(N-q)\mu$. However, we find an $N$-dependent number of 
  additional scars in the middle of the spectrum, i.e., at $-N\mu/2$.
For $N = 12$,
we find 15 scar states at $-N\mu/2$, and for $N = 16$ there are 19 scars also at $-N\mu/2$, with both numbers including one U(1) scar.
Because of the degeneracy they cannot be separated from the $U(1)$
scar state.  
As an example, we present in Fig.~\ref{fig:complex_N=12_eigval_scatter} (middle) a scatter plot of the eigenvalues of the
complex SYK model Eq.~(\ref{eq:sykhc}) with linear jump operators for $N=12$ and $\mu = 0.1$.

\section{Characterization of Lindblad scars by operator size}\label{sec:operator_size}
\label{sec:size}

So far we have not analyzed the properties of the eigenstates of scar states, which we expect to behave qualitatively differently from the surrounding quantum chaotic eigenstates.
In this section, we address this issue by showing that the operator size, which is a popular indicator of quantum chaos~\cite{Roberts_2018_Operator_growth, qi_2019_operatorsizegrowth, qi2019JHEP,schuster_2022_mbq_teleport, schuster_2022_operatorgrowthopenquantum,bhattacharjee2023JHEP,mori_2024,richter2025PRE},
provides a sharp distinction between the Lindblad scars introduced earlier and the rest of the eigenstates, which in the SYK model, are expected to be quantum chaotic.

In a fermionic system, the operator size is a measure of the (average) number of Majorana operators in an expansion of the operator in the basis of Majorana strings. A Majorana string
of size $p$ is given by
\begin{equation}
  \Gamma_{\vec n}=2^{p/2} \psi_{n_1} \cdots \psi_{n_p},
  \end{equation}
with $n_1\leq n_2\leq \dots \leq n_p$. 
It satisfies the orthogonality relation
${\Tr}\left[\Gamma_{\vec{n}} \Gamma_{\vec{m}}\right]=\delta_{\vec{n}\vec{m}}$.
Hence, the set of all $2^N$ possible strings forms a basis of the space of operators on the Hilbert space $\mathcal{H}_1$ and any operator can be expressed as a linear combination of these basis operators.
The size operator $\mathcal{S}$ can be defined as
\be
\label{eq:def_opsize}
\mathcal{S}(\hat{\mathcal{O}}) = \sum_{k=1}^N [\hat{\mathcal{O}},\psi_k ]_\pm \psi_k=\frac{N}{2}\hat{\mathcal{O}}\pm\sum_{k=1}^N \psi_k \hat{\mathcal{O}}\psi_k,
\ee   
where we choose the anti-commutator when the number of fermions in $\mathcal{O}$ is odd. We immediately see that the vectorized $\mathcal{S}$,
\begin{equation}\label{eq:size}
	\mathcal{S}=\frac{N}{2} - \i \sum_i \chi_i^L \chi_i^R,
\end{equation}
    satisfies $\mathcal{S}= -H_I/\mu$
    for $H_I$ given by Eq.~(\ref{eq:HI}) with our choice of jump operators, $L_i=\psi_i$. Hence, it follows that condition (ii') is equivalent to $\hat{\mathcal{O}}$ being an eigenstate of the size superoperator.
More precisely, we have that
\be
\mathcal{S}(\Gamma_{\vec n}) = p \Gamma_{\vec n},
\ee
i.e., the Majorana strings are the eigenvectors of the size superoperator with eigenvalue $p$ (the length of vector $\vec n$).

A generic operator $\hat{\mathcal{O}}$ is
not an eigenstate of the size operator. In this case, we can define its size as
\be
\langle{\mathcal S}\rangle_{\hat{\mathcal{O}}}=\frac{\Tr\left[\hat{\mathcal{O}}^{\dagger} \mathcal{S}(\hat{\mathcal{O}})\right] }
{\Tr \left[\hat{\mathcal{O}}^{\dagger} \hat{\mathcal{O}}\right]}.
\ee

\begin{figure*}[t]
	\includegraphics[width=0.49\textwidth]{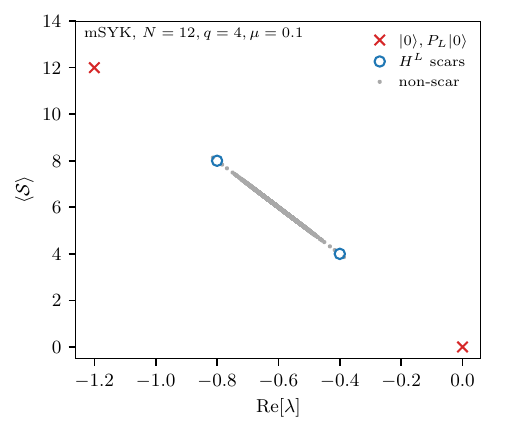}
   \includegraphics[width=0.49\textwidth]{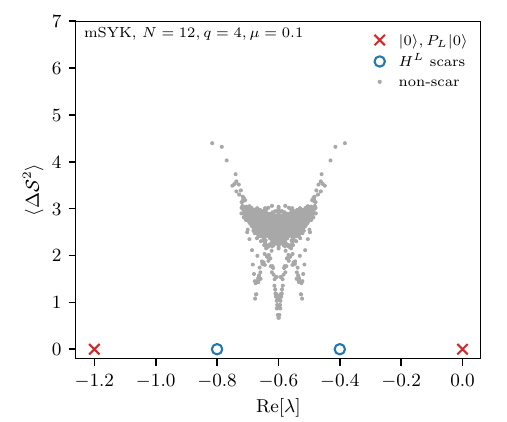}
	\caption{
	  Operator size Eq.~(\ref{eq:size}) of the eigenstates for  one realization of the vectorized Liouvillian of the Majorana SYK model Eq.~(\ref{eq:Maj_SYK_single_Hamil})	for $N = 12$, $q=4$, and $\mu=0.1$ as a function of the real part of the eigenvalues
          $\Re[\lambda]$. Scar states are denoted by red crosses and blue circles  and non-scar states by gray dots.
	Left: Operator size average, Eq.~(\ref{eq:size_vec}). There is a simple linear relation between the size of the eigenstates and the real part of the corresponding eigenvalues, as expected.
	Right: Operator size variance,  
	$\Delta \mathcal{S}^2 = \langle\mathcal{S}^2\rangle - \langle\mathcal{S}\rangle^2$. Only the scar states have a distinct vanishing variance since they are shared eigenstates of the Liouvillian and the size operator.
        }
	\label{fig:majorana_N=12_size}
\end{figure*}

Denoting the normalized right eigenstates of the vectorized Liouvillian by $|\alpha\rangle$, the size of this state is given by 
\begin{equation}\label{eq:size_vec}
	\langle {\mathcal S}\rangle =N / 2 - \sum_{i=1}^{N}\expval{\chi_i^L \chi_i^R}{\alpha}.
\end{equation}
Here and below we use the notation  $\langle \cdot \rangle $ for
the expectation value  $\langle\alpha | \cdot |\alpha \rangle $ of state $|\alpha\rangle$.

In Fig.~\ref{fig:majorana_N=12_size}, we depict results for the average and variance of the size Eq.~(\ref{eq:size_vec}) for the Liouvillian of the $q=4$ SYK Hamiltonian Eq.~(\ref{eq:sykhc}) with linear jump operators and $\mu=0.1$.
Since the real part of the eigenvalues $\lambda$ of $\mathcal{L}$ is equal to the expectation value
of $H_I$ we find that the size is a linear function of
$\Re[\lambda]$, see Fig.~\ref{fig:majorana_N=12_size} (left). Therefore, the size cannot be used to distinguish scar from non-scar states.
However, since scars are shared eigenstates of the Liouvillian and the size operator, the variance of the scar states vanishes, while for the non-scar states it is positive (see Fig.~\ref{fig:majorana_N=12_size}, right), and can thus be used to distinguish scar states from non-scar states.
Their number agrees with the discussion in Sec.~\ref{sec:operator scars}.

\begin{figure*}[t]
	\includegraphics[width=0.49\textwidth]{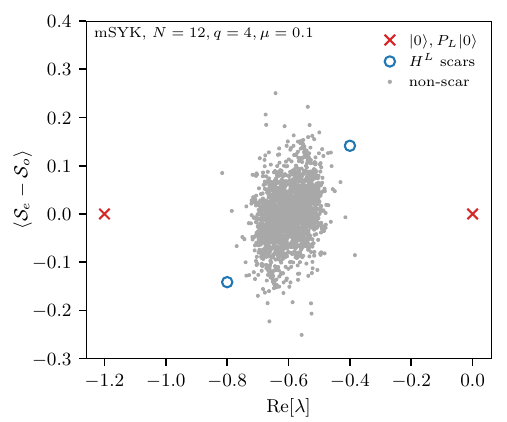}
	\includegraphics[width=0.49\textwidth]{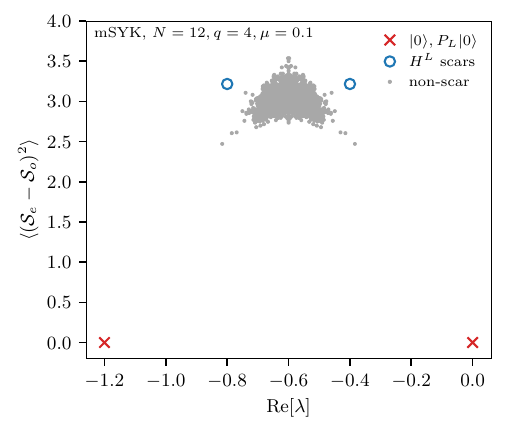}
	\caption{Expectation values of $\mathcal{S}_e- \mathcal{S}_o$, Eq.~(\ref{eq:sizeo}) (left), and $(\mathcal{S}_e- \mathcal{S}_o)^2$, Eq.~(\ref{eq:sizeovar}) (right), for one realization of the vectorized Liouvillian of the Majorana SYK model Eq.~(\ref{eq:Maj_SYK_single_Hamil})	for $N = 12$, $q=4$, and $\mu=0.1$ as a function of the real part of the eigenvalues $\Re[\lambda]$. Scar states are denoted by red crosses and blue circles and non-scar states by gray dots.
	Only the parity scars have vanishing moments of $\mathcal{S}_e-\mathcal{S}_o$ because they are simultaneous eigenvalues of $\mathcal{S}_e$ and $\mathcal{S}_o$.
	}
	\label{fig:syk-eo}
\end{figure*}
	
 \begin{figure*}[t]
	\includegraphics[width=0.49\textwidth]{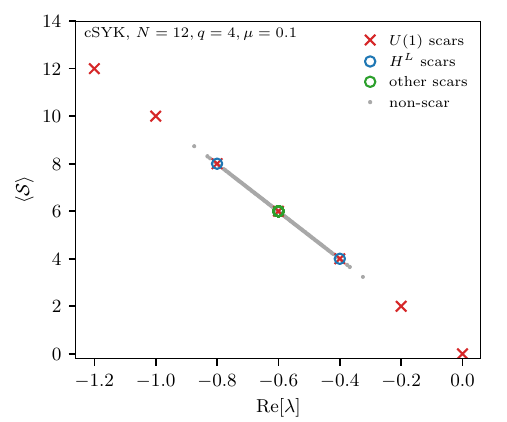}
	\includegraphics[width=0.49\textwidth]{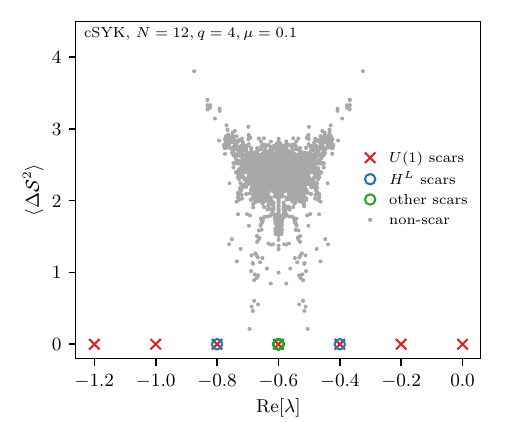}
	\caption{
	Operator size Eq.~(\ref{eq:size}) of the eigenstates of one realization of the vectorized Liouvillian of the complex SYK model Eq.~(\ref{eq:sykhc}) for $N = 12$, $q=4$, and $\mu=0.1$ as a function of the real part of the eigenvalues $\Re[\lambda]$. Scar states are denoted by red crosses and blue and green circles and non-scar states by gray dots.
	Left: Operator size average, Eq.~(\ref{eq:size_vec}).
	Right: Operator size variance,  
	$\Delta \mathcal{S}^2 = \langle\mathcal{S}^2\rangle - \langle\mathcal{S}\rangle^2$. 
	As in the Majorana case, see Fig.~\ref{fig:majorana_N=12_size}, the mean of the size depends linearly on $\Re[\lambda]$ and the variance of the sizes vanishes only for scars states.}
	\label{fig:complex_N=12_size}
\end{figure*} 

\begin{figure*}[t]
	\includegraphics[width=0.49\textwidth]{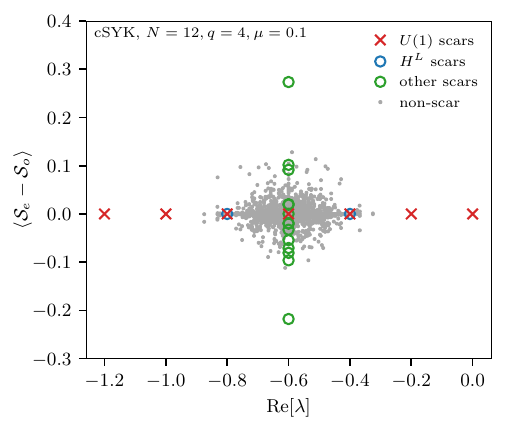}
	\includegraphics[width=0.49\textwidth]{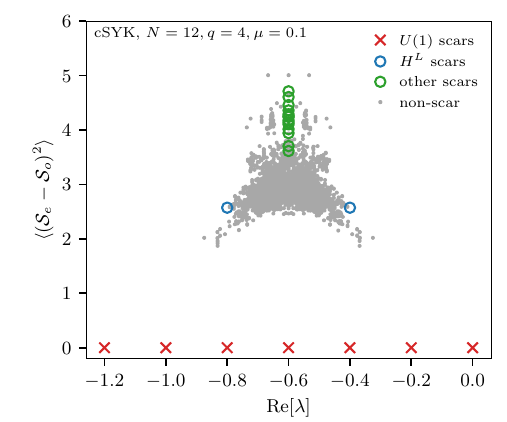}
	\caption{
 Expectation values of $\mathcal{S}_e- \mathcal{S}_o$
            , Eq.~(\ref{eq:sizeo}) (left),
and $(\mathcal{S}_e- \mathcal{S}_o)^2$, Eq.~(\ref{eq:sizeovar}) (right),
 for  one realization of the vectorized Liouvillian of the complex SYK model Eq.~(\ref{eq:sykhc}) for $N = 12$, $q=4$, and $\mu=0.1$, as a function of the real part of the eigenvalues $\Re[\lambda]$. Scar states are denoted by red crosses and blue and green circles and non-scar states by gray dots.
		Only the $N/2+1$ $U(1)$ scars are simultaneous eigenstates of $\mathcal{S}_e$ and $\mathcal{S}_o$ and hence have vanishing moments of $\mathcal{S}_e-\mathcal{S}_o$.
	}
		\label{fig:csyk-eo}
	\end{figure*}

To characterize scar states further, we can eliminate the dependence of the operator size on the real part of the eigenvalues by separating the size contribution into different parts, e.g., odd and even Majoranas. Specifically, in the complex SYK model, we define
\begin{align}
	\mathcal{S}_{o} &=\frac N4- \i \sum_{i=1}^{N/2} \chi_{2i-1}^L \chi_{2i-1}^R,
	\\
	\mathcal{S}_e &= \frac{N}{4}-\i \sum_{i=1}^{N/2}\chi_{2i}^L \chi_{2i}^R.
\end{align}
To eliminate the dependence on the total size,
we will focus on the observable $\mathcal{S}_e-\mathcal{S}_o$ in the rest of the section.
We consider the expectation values
\be \label{eq:sizeo}
\langle\mathcal{S}_e-\mathcal{S}_o\rangle 
\ee
and
\be \label{eq:sizeovar}
\langle(\mathcal{S}_e-\mathcal{S}_o)^2\rangle.
\ee

In the Majorana SYK case, see Fig.~\ref{fig:syk-eo}, we observe that the average and the
second moment
of ${\mathcal S}_e-{\mathcal S}_o$ are qualitatively different for scars and non-scar states. Moreover, the TFD scar $|0\rangle$ is the fermionic vacuum and hence has the same number of even and odd Majoranas, namely, zero. Likewise, the scar $P^L|0\rangle$ is the completely filled state and has the maximal number $N/2$ of even and odd Majoranas. Accordingly, both scars are eigenstates of both $\mathcal{S}_e$ and $\mathcal{S}_o$ with the same eigenvalue and the expectation values in Eqs.~(\ref{eq:sizeo}) and (\ref{eq:sizeovar}) vanish. On the other hand,
$H^L$ is a sum of terms that do not have a fixed number of even (or odd) Majoranas; indeed, there are terms with $0$, $1,\dots,q$ even (or odd) Majoranas. Therefore, the two $H^L$ scars are not eigenstates of $\mathcal{S}_e$ and $\mathcal{S}_o$ and the expectation values in Eqs.~(\ref{eq:sizeo}) and (\ref{eq:sizeovar}) are finite.

We now turn to the complex SYK case, Eq.~(\ref{eq:sykhc}). For the sake of completeness, we first study the operator size and variance without separating the size contribution of the different parts. We observe, see Fig.~\ref{fig:complex_N=12_size}, a similar pattern as in the Majorana case, although with a much richer scar spectrum, which will be discussed shortly. The operator size is still linear in  the real part of the eigenvalues which make it unsuitable for a general characterization of scars. Likewise, scar states size have zero variance because the Liouvillian and the size operator have common eigenstates. Therefore, from now on, we follow the route of the Majorana fermions, and focus on the more interesting case of computing the size of odd and even fermion separately. We refer to the previous discussion on Majorana SYK for further details. 
   
The results depicted in Fig.~\ref{fig:csyk-eo} for ${\mathcal S}_e - {\mathcal S}_o$ permit a more detailed study of the richer pattern of many-body scars observed for complex fermions. The predicted $U(1)$ and $H^L$ scars have zero average $\langle{\mathcal S}_e - {\mathcal S}_o\rangle$, Eq.~(\ref{eq:sizeo}), which is the theoretical prediction. Indeed, the $U(1)$ scars are constructed from the operators $\hat{\mathcal{N}_p}$, Eq.~(\ref{eq:tpl}), which always have the same number of even and odd Majoranas. Similarly, the projection in $\hat{H}_c$ ensures\footnote{We can show this explicitly for $q=2$ and conjecture it to also hold for larger $q$ based on the numerical results.} that, when expressed in terms of Majorana fermions,  for every term in $\widehat H_c$ 
  with a given value of $\mathcal{S}_e-\mathcal{S}_o$, there exists another term with the same coupling but opposite $\mathcal{S}_e-\mathcal{S}_o$, and
  $\mathcal{S}_e(\widehat H_c)-\mathcal{S}_o(\widehat H_c) $ vanishes.
Moreover, $\hat{\mathcal{N}}_p$ is a simultaneous eigenstate of both $\mathcal{S}_e$ and $\mathcal{S}_o$ with the same eigenvalue and, hence, also the variance, Eq.~(\ref{eq:sizeovar}), vanishes for the $U(1)$ scars. On the other hand, while $\hat{H}_c$ has $\langle{\mathcal S}_e - {\mathcal S}_o\rangle=0$ on average, it is not an eigenstate of $\mathcal{S}_e$ and $\mathcal{S}_o$ and therefore the variance is finite.

Surprisingly, we have found 15\footnote{The U(1) scar at the
center of the spectrum is completely mixed among these states and cannot
be identified separately. However, we can construct it analytically as
$\mathcal{N}^L_{N/4} |0\rangle$.}  
 scars at $\lambda = -0.6=-N\mu/2$ that we have labeled as ``other scars''. These states are scars because they are eigenstates of $-iH_0$ and $H_I$ with  eigenvalue $-N\mu/2$. However, we could not identify to what symmetry of SYK, if any, they are related to. Interestingly, these  scars have a varied range of operator size and variance
that reinforces the idea that scars are promising resources for quantum information applications. We note that the analytical $U(1)$ scar at the center of the spectrum is also contained among the ``other scars'', and is thus plotted twice.
Without a full characterization of these degenerate scars we cannot conclude whether this degeneracy with the $U(1)$ scar is
 based on symmetry, but
there is a possibility that they are connected with a symmetry of the SYK model, or a feature of the jump operators. Therefore, they need not be restricted to this model, although we do not observe them in the XXZ spin chain to be studied in Sec.~\ref{sec:spin}.  
   
\begin{figure}[t]
	\includegraphics[width=\columnwidth]{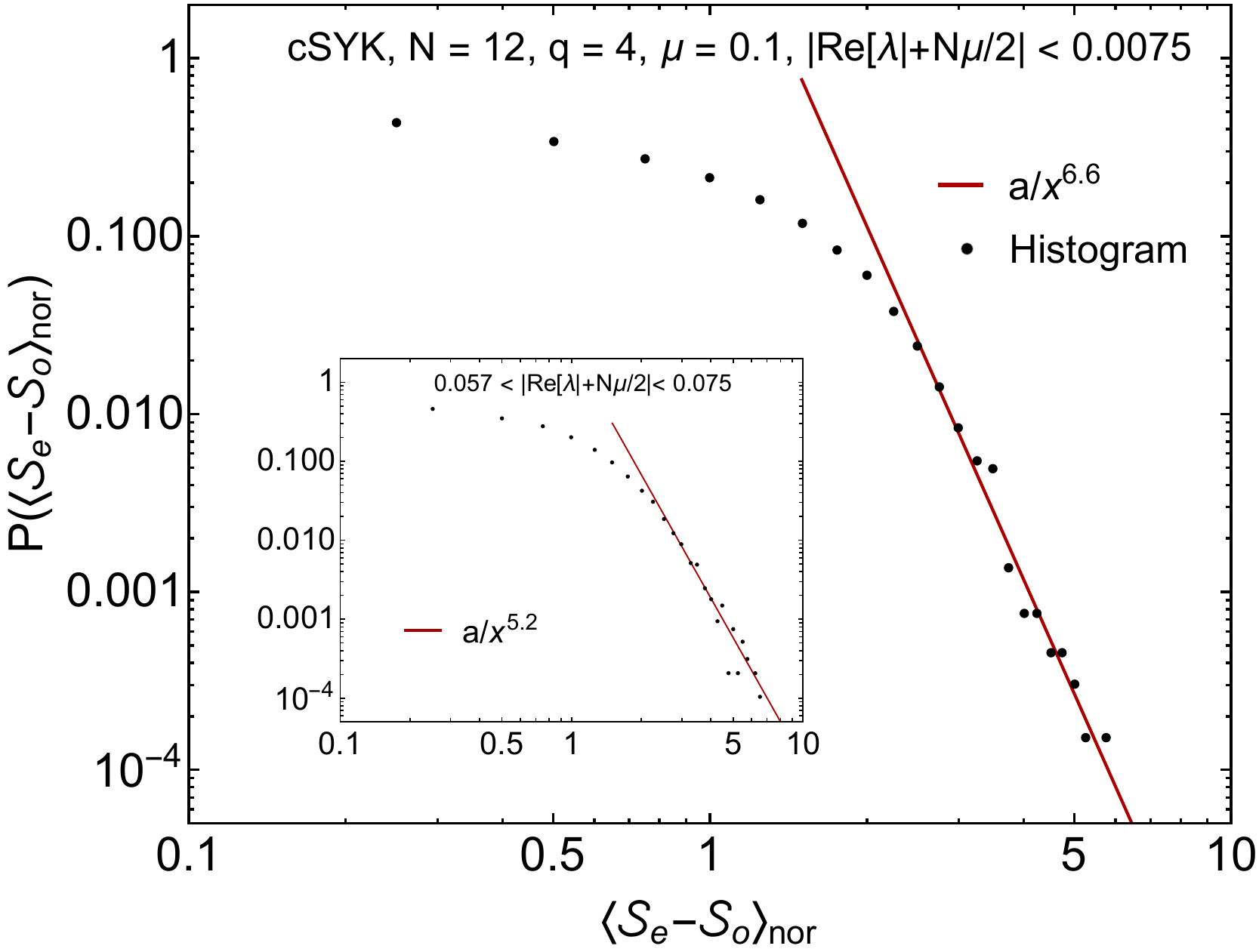}
	\caption{Log-log plot of the histogram of the normalized expectation values
          $\langle{\mathcal S}_{e} - {\mathcal S}_{o}\rangle$ (black points) for the non-scar states of the complex SYK model with $N=12$, $q=4$, and $\mu =0.1$ (for 200 realizations).
          The distribution is computed for a small window around the center of the spectrum (main)
          and a window away from the center (inset) with values as indicated in the figure.
          A fit by the power law $a / x^b$ (red),  with the parameters $a$ and $b$ depending on the window of eigenvalues,  provides a good description of the asymptotic decay of the distribution. The observed deviations from a Gaussian behavior suggest that ETH in dissipative quantum chaos may be qualitatively different from that in Hermitian quantum chaos \cite{srednicki1994,deutsch1991}. 
		  }
	\label{fig:eohis}
\end{figure}

Finally, we address the full distribution of the operator size $\langle\mathcal{S}_e-\mathcal{S}_o\rangle$. For  states with a nonvanishing expectation value
(among other scar states but also many other states)
it can be used as a powerful probe of the applicability and relevance of the ETH in many-body quantum dissipative systems.
According to the Hermitian ETH \cite{srednicki1994,deutsch1991}, we would expect
a Gaussian distribution of $\langle {\mathcal S}_{e}-{\mathcal S}_{o}\rangle$ with a mean and a  width that show a smooth variation as a function of the energy 
 (i.e., an
  energy dependence on a scale that is much larger than the scale of the fluctuations)~\cite{draayer1977strength,draayer1977strengthII,Verbaarschot:1984phu,verbaarschot1985distribution}.
 For non-Hermitian operators, the spectrum is
 two-dimensional, and the size operator can have a nontrivial smooth behavior as a function
 of  both
 the real and imaginary parts of the eigenvalues.
 This can be  shown by calculating
 the mean and standard deviation of the nonzero size differences
 as a function of the eigenvalue using a small
bin size and fitting them by  smooth functions, $M(\lambda)$ and $\sigma(\lambda)$,
in this order.
 The normalized size difference for a state with eigenvalue $\lambda$
is defined by
\be
\langle {\mathcal S}_{e}-{\mathcal S}_{o}\rangle_{\rm nor}
=\frac{\langle {\mathcal S}_{e}-{\mathcal S}_{o}\rangle -M(\lambda)}
     {\sigma(\lambda)}.
     \ee
      As already suggested by Fig. \ref{fig:csyk-eo} (left), their dependence
 on  the real and imaginary parts
 of the eigenvalues is relatively weak.  After normalization to zero mean and unit variance
 we can combine results to form a group of nearby bins to increase statistics.

 After performing this normalization procedure, we find the distribution shown in Fig.~\ref{fig:eohis} (black dots) for $|{\rm Re}[\lambda] +N\mu/2| <0.0075$ (main) and
$0.057<|{\rm Re}[\lambda] +N\mu/2| <0.075$ (inset). In both cases, it has a distinct power-law  tail (red curve) with a power that decreases
away from the center.
 Note that on average, the value of 
 $\langle {\mathcal S}_{e}-{\mathcal S}_{o}\rangle$
vanishes
for about 62$\%$ of the states.

While a Gaussian distribution is the broadly accepted expectation for Hermitian systems, the reach and details of the ETH for non-Hermitian systems is still under debate \cite{hamazaki2022arXiv,roy2023,mao2024,cipolloni2024,almeida2025}.
Because the non-stationarity
  of  the normalized  distribution of $\langle {\mathcal S}_{e}-{\mathcal S}_{o}\rangle_{\rm nor} $
  is negligible for the energy windows for which the distribution was calculated
 we can exclude this as  the cause of the non-Gaussianity.
Therefore, these results point to qualitative differences with respect to the Hermitian case that deserve further consideration.

\section{Characterization of Lindblad scars by entanglement entropy}
\label{sec:EE}

The analytical properties and fixed operator sizes of the U(1) scars imply a rather simple operator complexity. A natural conjecture stepped further is that the EE of the scars are subvolume given their simple operator structure. We will see that for Lindblad scars the picture is more complicated because the EE depends sensitively on the choice of the partition.
A detailed characterization of the entanglement of the Lindblad scars is important not only for
the understanding of the scar itself but also to give insights into quantum many-body effects such as quantum chaos and information scrambling.
  
\begin{figure*}[t]
	\includegraphics[width=0.49\textwidth]{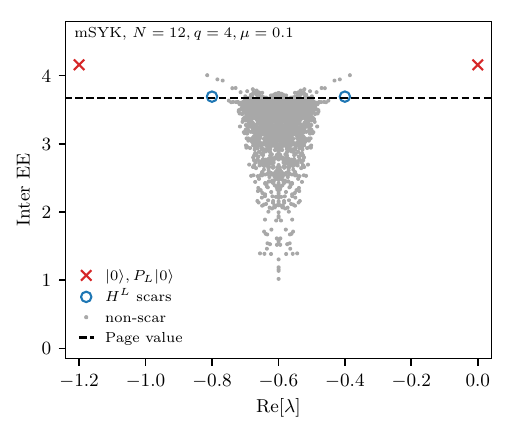}
	\includegraphics[width=0.49\textwidth]{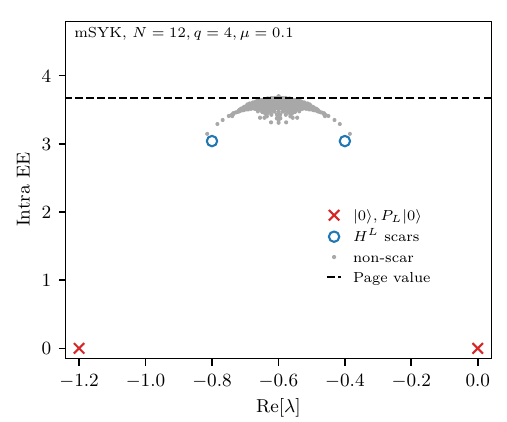}
	\caption{Entanglement entropy Eq.~(\ref{eq:ee}) of the eigenstates of one realization of the vectorized Liouvillian of the Majorana SYK model Eq.~(\ref{eq:Maj_SYK_single_Hamil}) for $N=12$, $q=4$, and $\mu=0.1$, using two different partitions of the degrees of freedom. 
		Left: Intersite partition corresponding to the original bipartition of $2N$ Majoranas into $N$ $L$ Majoranas and $N$ $R$ Majoranas.
		Right: Intrasite partition corresponding to a non-contiguous split, see the main text for details. 
		The EE of scars states is sensitive to both the choice of partition and the type of scars.
  The TFD scar states (red crosses) are clearly differentiated from non-scar states (gray dots) but the $H^L$ scars (blue circles) have an EE that is within the range of the non-scar states.}

	\label{majorana_N=12_EE}
\end{figure*}

\begin{figure*}[t]
	\includegraphics[width=0.49\textwidth]{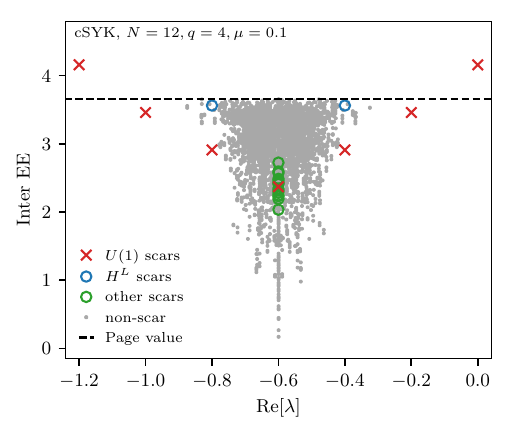}
	\includegraphics[width=0.49\textwidth]{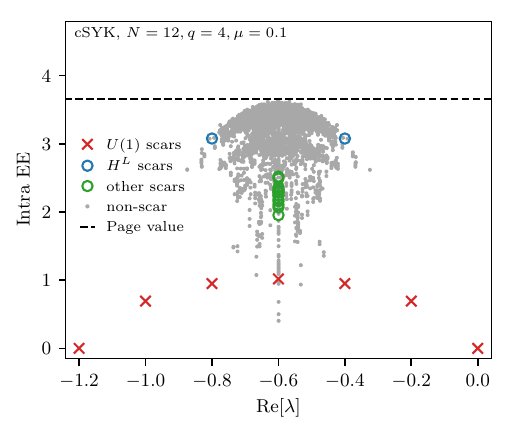}
	\caption{
		Entanglement entropy Eq.~(\ref{eq:ee}) of the eigenstates of one realization of the vectorized Liouvillian of the complex SYK model Eq.~(\ref{eq:sykhc}) for $N=12$, $q=4$, and $\mu=0.1$, using two different partitions of the degrees of freedom. 
		Left: Intersite partition corresponding to the original bipartition of $2N$ Majoranas into $N$ $L$ Majoranas and $N$ $R$ Majoranas.
		Right: Intrasite partition corresponding to a non-contiguous split, see the main text for details.
		The EE of scars states (red crosses and green and blue circles) has a broad range of values, from above the Page value typical of a highly entangled random state down to zero entanglement. As in the Majorana case, the EE is sensitive to both the choice of partition and the type of scars. The $U(1)$ scars (red crosses) are only present in the complex SYK model. We also observe that the EE of each of the 15 degenerate scars that we cannot relate to a symmetry (green circles) is different and does not overlap with the rest of scars, except with one $U(1)$ scar with the same eigenvalue.}
	\label{fig:complex_N=12_EE}
\end{figure*}  
 
The EE of eigenstate $|k\rangle$ of $\mathcal{L}$ is defined as
\begin{equation}\label{eq:ee}
	S^{(k)}_\Omega=-\Tr[\rho^{(k)}_\Omega\log \rho^{(k)}_\Omega], \quad 
	\rho^{(k)}_\Omega=\Tr_{\overline{\Omega}}|k\rangle\langle k|,
\end{equation}
where $\Omega$ is a subset of the degrees of freedom in the doubled Hilbert space $\mathcal{H}_2$ and $\Tr_{\overline{\Omega}}$ denotes the partial trace over the complementary degrees of freedom. In practice, to compute $S^{(k)}_\Omega$ we bipartition the doubled Hilbert space as $\mathcal{H}_2=\mathcal{H}_\Omega\otimes \mathcal{H}_{\overline{\Omega}}$ and perform the Schmidt decomposition of $|k\rangle$,
\begin{equation}
	|k\rangle=
	\sum_{ab}K_{ab}^{(k)}
	|a\rangle_\Omega|b\rangle_{\overline{\Omega}},
\label{sumK}
\end{equation}
where $|a\rangle_\Omega$ and $|b\rangle_{\overline{\Omega}}$ are bases of $\mathcal{H}_\Omega$ and $\mathcal{H}_{\overline{\Omega}}$, respectively. With $\sigma_\ell$ the singular values of the matrix $K^{(k)}_{ab}$, $\ell=1,\dots,\min\{\mathrm{dim}(\Omega),\mathrm{dim}(\overline{\Omega})\}$, the EE is given by
\begin{equation}
	S^{(k)}_\Omega=-\sum_\ell \sigma_\ell^2\log \sigma_\ell^2.
\label{entan}
\end{equation}
We considered the following two bipartitions of the degrees of freedom, which lead to qualitatively different entanglement patterns:

\begin{itemize}
\item The \textit{intersite} partition corresponds to the original bipartition of the $2N$ Majoranas into $N$ $L$  Majoranas and $N$ $R$ Majoranas, $\mathcal{H}_2=\mathcal{H}^L\otimes \mathcal{H}^R$ (i.e., $\Omega=L$).

\item For the \textit{intrasite} partition, we split the $L$ Majoranas into two subsets, $\mathcal{H}^L=\mathcal{H}^{L}_A\otimes\mathcal{H}^{L}_B$, where $A$ corresponds to the first $N/2$ Majoranas and $B$ to the last $N/2$ Majoranas,
  and similarly for $\mathcal{H}^R$. The Schmidt decomposition is then performed with respect to the bipartition $\mathcal{H}_2=\mathcal{H}_A\otimes \mathcal{H}_B$ (i.e., $\Omega=A$),   where  $\mathcal{H}_A=\mathcal{H}^{L}_A\otimes\mathcal{H}^{R}_A$ and
$\mathcal{H}_B=\mathcal{H}^{L}_B\otimes\mathcal{H}^{R}_B$. 
  For a state $|k\rangle $, the decomposition \eref{sumK} is expressed as
  \be
  |k\rangle = \sum_{i_1i_2; i_3i_4} K_{i_1i_2; i_3i_4}^{(k)}  |i_1i_3\rangle |i_2i_4\rangle,
\label{intrasite}
  \ee
  where  $|i_1i_3\rangle\in \mathcal{H}^{L}_A\otimes\mathcal{H}^{L}_B$ and
  $|i_2i_4\rangle\in \mathcal{H}^{R}_A\otimes\mathcal{H}^{R}_B$
  (see Appendix \ref{appA} for more details).
  
    \end{itemize}

  The choice of partition, and hence the calculation of the EE, has implicit a choice of basis (i.e., of representation of the fermion operators and vectorization scheme) which can significantly affect the values of the EE. In particular, the vectorization of Eq.~(\ref{eq:fermions1}) introduces an explicit coupling between the two copies of the Hilbert space and therefore affects the intersite EE. To avoid the extra spurious entanglement, we choose the vectorization of Eq.~(\ref{eq:vectorization_Maj}); ultimately, this choice is arbitrary and the most appropriate one must be done on a case-by-case basis.

In Figs.~\ref{majorana_N=12_EE} and \ref{fig:complex_N=12_EE}, we show the EE for the dissipative Majorana and complex SYK models, respectively. 
The  EE of many of the non-scar states is close to the Page value (the value for a typical random state),
\begin{equation}
	S_{\mathrm{Page}}=
	\sum_{j=D+1}^{D^2}\frac{1}{j} 
	-\frac{D - 1}{2D},
\end{equation}
with $D=2^{N/2}$ the Hilbert space dimension.
On the other hand, as expected, the scar states have a different entanglement pattern for
the two partitions. This is particularly visible for the complex SYK model because of the higher number of scars induced by the $U(1)$ symmetry. In particular, we see that the scars with high entanglement in one partition have low entanglement in the other. 
Consider, for example, the states $|0\rangle$ and $P^L|0\rangle$. Since these states correspond to the infinite-temperature TFD and its parity reflection, they have maximal entanglement between the $L$ and $R$ spaces; correspondingly, the intersite EE reaches the maximum value $\log D=\frac N2\log2$. On the other hand, the intrasite EE vanishes for
this state (see Appendix \ref{appA}).
The other $U(1)$ scars have larger intrasite EE than the TFD or its parity reflection,
but it is still noticeably less than that of the bulk of the non-scar states.
Since the intrasite EE we consider here coincides with the \textit{operator EE}~\cite{prosen2007operator,zhou2017operator,wellnitz2022rise}, this low EE for this partition is an indication of a simple operator complexity. This is no surprise if one looks explicitly into the defining expressions [see Eq.~(\ref{eq:tpl})]. By contrast, the definition of the $H_L$ scar does not point to a simple operator complexity. Indeed, its EE is closer to that of non-scar states and in all cases does not vanish. 
\section{Lindblad scars in spin chains}
\label{sec:spin}

The existence and characterization of Lindblad scars is not exclusive to the dissipative SYK model discussed previously. In this section, we show that similar findings also occur in a dissipative random spin-$1/2$ XXZ chain whose Hamiltonian is given by
\begin{equation} \label{eq:xxz_hamiltonian}
\hat{H}_{S} = - J \sum_{i=1}^{N} (\hat{X}_i \hat{X}_{i+1} + \hat{Y}_i \hat{Y}_{i+1} + \Delta \hat{Z}_i \hat{Z}_{i+1}) - \sum_{i=1}^N h_i \hat{Z}_i,
\end{equation}
where $\hat{X}_i$, $\hat{Y}_i$, and $\hat{Z}_i$ are the standard Pauli matrices on site $i$, we have assumed periodic boundary conditions ($\hat{X}_{N+1}\equiv\hat{X}_1$, etc.), $h_i \in [-h,h]$ are random fields and $N$ is the number of sites. When $h/J$ is sufficiently small (large), the model is quantum chaotic (many-body localized)
\cite{luitz2015}. We set $h/J=0.5$ and $\Delta=1.1$ so that the system is in the quantum chaotic phase.
The XXZ Hamiltonian also has a $U(1)$ symmetry $[\hat{H}_{S},\sum_{i=1}^N \hat{Z}_i] = 0$.
We choose the jump operators to be $\hat{L}_{i} = \hat{X}_{i}$ so that they  break the $U(1)$ symmetry in the Liouvillian.
For our calculations we choose $\mu =0.1$ which corresponds to a much
weaker dissipation than in the SYK and complex SYK cases.

\begin{figure*}[t]
	\includegraphics[width=0.49\linewidth]{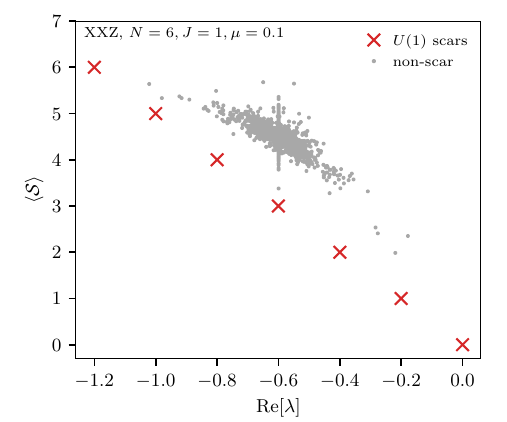}
	\includegraphics[width=0.49\linewidth]{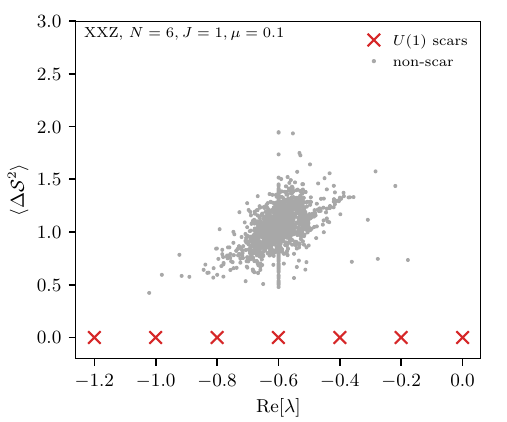}
\vspace*{0.5cm}	\includegraphics[width=0.49\linewidth]{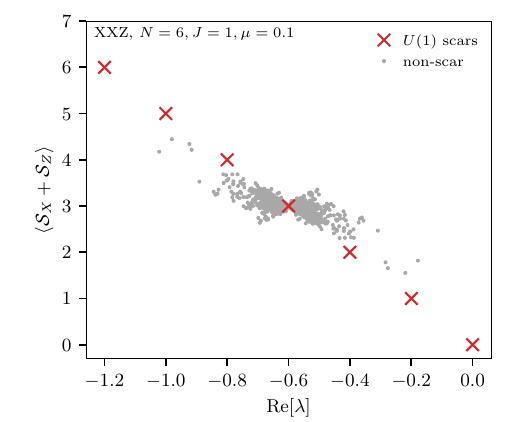}\hspace*{0.1cm}\vspace*{-0.6cm}
	\includegraphics[width=0.49\linewidth]{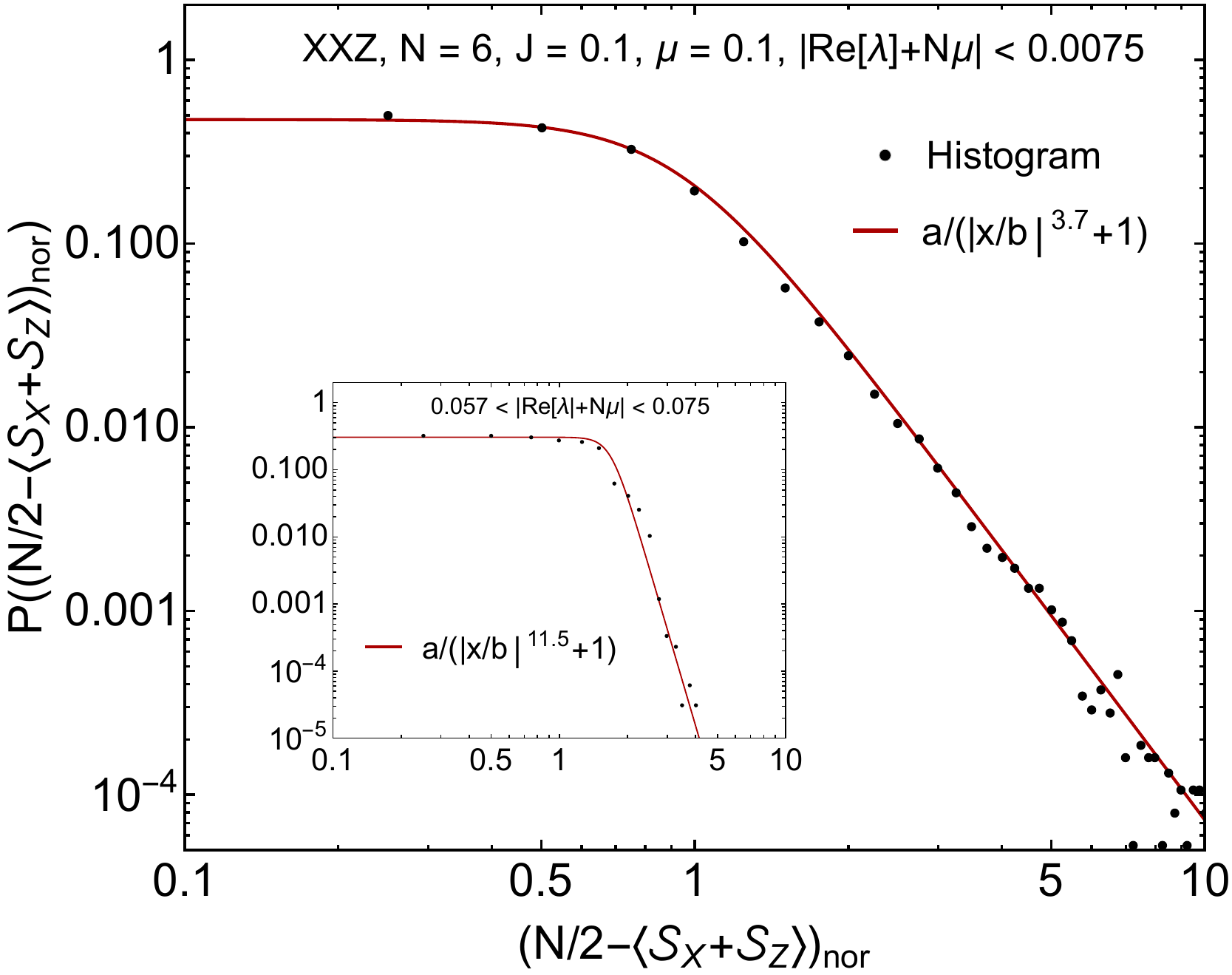}
	\caption{
	  Upper left: Operator size Eq.~(\ref{eq:sspin}) of scar states (red crosses) and non-scar states (gray dots) for one realization of the $N=6$
          dissipative XXZ model Eq.~(\ref{eq:xxz_hamiltonian}).
	Upper right: Operator size variance $\Delta \mathcal{S}^2 = \langle\mathcal{S}^2\rangle - \langle\mathcal{S}\rangle^2$. Lindblad scar states have vanishing size variance since they are shared eigenstates of the Liouvillian and size  operator.
	Lower left: Operator size for Pauli $X$ and $Z$ operators, Eqs.~(\ref{eq:sspinx}) and (\ref{eq:sspinz}). Scar states have size $-{\rm Re}[\lambda]/2\mu$ since they only have components from $Z$ operators so this observable can be used to characterize scar state in this model. 
	Lower right: Log-log plot of histogram of the normalized operator size for Pauli $X$ and $Z$ operators for non-scar states. We observe a similar power-law decay as in the SYK case that suggests deviations from the ETH for dissipative quantum chaotic systems.}
	\label{fig:spin_size}
\end{figure*}

The vectorization of the Liouvillian is given by \cite{sa2019csr,sa2022a,kawabata2023symmetry}:
\begin{equation}
\begin{aligned}
\mathcal{L}= & -\i\left(\hat{H}_S \otimes \id-\id \otimes \hat{H}_S^*\right) \\
& +\mu\sum_\alpha\left[\hat{L}_\alpha
  \otimes \hat{L}_\alpha^*-\frac{1}{2}\left(\hat{L}_\alpha^{\dagger} \hat{L}_\alpha \otimes \id
  + \id \otimes (\hat{L}_\alpha^\dagger \hat{L}_\alpha)^*\right)\right].
\end{aligned}
\end{equation}
Defining
\begin{align}
	&X^L_i=\hat{X}_i\otimes \id, \qquad
	X^R_i=\id\otimes\hat{X}^*_i, 
	\\
	&Y^L_i=\hat{Y}_i\otimes \id, \qquad
	Y^R_i=\id\otimes\hat{Y}^*_i, 
	\\
	&Z^L_i=\hat{Z}_i\otimes \id, \qquad
	Z^R_i=\id\otimes\hat{Z}^*_i, 
\end{align} 
we can equivalently write
\begin{equation}
\mathcal{L} = - \i \left(H_{S}^L - H_{S}^R\right) + H_I,
\end{equation}
with $H_S^L$ obtained from $\hat{H}_S$ by replacing $\hat{X}_i$ by $X_i^L$, etc., and
\begin{equation}
  H_I = \mu\sum_{i=1}^N X^L_i X^R_i-N\mu.
  \label{hispin}
\end{equation}
The vectorized Lindbladian  has the following discrete symmetries,
\begin{align}
    &[\prod_k Z_k^LZ_k^R, \mathcal{L}] =0,
    \label{S1}
    \\
    &\{ {\rm SWAP} \times \prod_k Z_k^L, \mathcal{L}+\mu N \} =0,
    \label{S2}
    \\
	&[  {\rm SWAP}  \times  K, \mathcal{L}]=0,
	\label{S3}
\end{align}
where SWAP is the operator that interchanges $L$ and $R$ and $K$  is complex conjugation.
    Because of Eq.~\eref{S2}, the eigenvalues of $ \mathcal{L}+\mu N$
    occur in pairs $\{ \lambda_k, -\lambda_k\}$,
    while Eq.~\eref{S3} implies that they appear as complex conjugate pairs if not real.
    
     Although the operator Eq.~\eref{hispin} resembles $H_I$ of the SYK and complex SYK models,
  its properties are significantly different. Its ground state is $2^N$ fold
  degenerate,
  while the ground state of $H_I$ in Eq. \eref{hi} is unique.
  The ground-state degeneracy of $H_I$ is lifted by a nonzero value of $J$,
  and the steady state of $-iH_0+H_I+N\mu$ is only well separated from the rest of the spectrum for
sufficiently  small values of $\lVert( H_I +N\mu)\rVert/\lVert H_0 \rVert$.
This is the case for our choice of $\mu = 0.1$ for which
$\lVert H_I+N\mu\rVert/\lVert H_0 \rVert \approx 0.05$ compared with 0.5 for the corresponding ratio
for
the complex SYK  (compare
the complex SYK spectrum and the XXZ spectrum in 
Fig. \ref{fig:complex_N=12_eigval_scatter}).
At this value of $\mu$ about $40\%$ of the eigenvalues of $\mathcal{L}+N\mu$ are purely imaginary.

Due to the $U(1)$ symmetry that is broken by the jump operators, we expect the presence of $U(1)$ scars. 
A straightforward calculation shows that the role of the $p$-tuple operator ${\hat{\mathcal{N}}}_p$ Eq.~(\ref{eq:tpl}) that generates the scars in the complex SYK model is played by 
\be 
{\mathcal M}_p=\sum_{k_1\neq k_2\neq\cdots\neq k_p} Z_{k_1} Z_{k_2} \cdots Z_{k_p}.
\ee
Since it verifies the two conditions (i) and (ii'), the scars in the dissipative spin-chain $U(1)$ are ${\mathcal M}_p^L|0 \rangle$ with $p=0,\dots,N/2$,  whose eigenvalues are $-2p\mu$:
\begin{equation}
	\mathcal{L}\mathcal{M}_p^L|0\rangle=-2p\mu \mathcal{M}_p^L|0\rangle.
\end{equation}
Unlike in the SYK model, $H^L|0\rangle$ is not a scar because the Hamiltonian has terms with a different number of Pauli matrices: one Pauli matrix in the magnetic field term and two Pauli matrices in the exchange term.  This is confirmed by a scatter
plot of the eigenvalues,  see Fig.~\ref{fig:complex_N=12_eigval_scatter} (left) where the
scar states are shown by the red crosses.

To identify and further characterize these scars, we computed the size and the EE. 
Paralleling the definition introduced in Sec.~\ref{sec:operator_size}, the size operator of spin systems is given by~\cite{Roberts_2018_Operator_growth, qi_2019_operatorsizegrowth, qi2019JHEP,schuster_2022_mbq_teleport, schuster_2022_operatorgrowthopenquantum,bhattacharjee2023JHEP,mori_2024,richter2025PRE}
\begin{align}
&\mathcal{S}(\hat{\mathcal{O}}) =-\frac{1}{4}\sum_{\{\hat{\Sigma}_i\}}\left(\hat{\Sigma}_i \hat{\mathcal{O}} \hat{\Sigma}_i^{\dagger}-\hat{\mathcal{O}}\right),
 \\
&\mathcal{S}_{\Sigma}(\hat{\mathcal{O}}) = \mathcal{S}(\hat{\mathcal{O}}) + \frac{1}{2} \sum_{i=1}^N (\hat{\Sigma}_i \hat{\mathcal{O}} \hat{\Sigma}^{\dagger}_i - \hat{\mathcal{O}} )
\end{align}
with $\hat{\Sigma}_i \in\left\{\hat{X}_i, \hat{Y}_i, \hat{Z}_i\right\}$. Here, $\mathcal{S}$ measures the number of sites in which an operator is one of the Pauli matrices, while $\mathcal{S}_{\Sigma}$ measures the number occurrences of a specific Pauli operator $\Sigma$ only.
After vectorization, they can be written as
\begin{align}
\label{eq:sspin}
\mathcal{S} &= \frac{3N}{4} - \frac{1}{4}\sum_{\Sigma_i} \Sigma^L_i \Sigma^R_i, 
\\
\label{eq:sspinx}
\mathcal{S}_{X} &= \frac{N}{4} - \frac{1}{4}\sum_{i}^N(-X_i^L X_i^R + Y_i^L Y_i^R  + Z_i^L Z_i^R ),  
\\
\label{eq:sspiny}
\mathcal{S}_{Y} &= \frac{N}{4} - \frac{1}{4}\sum_{i}^N(X_i^L X_i^R - Y_i^L Y_i^R  + Z_i^L Z_i^R ),  
\\
\label{eq:sspinz}
\mathcal{S}_{Z} &= \frac{N}{4} - \frac{1}{4}\sum_{i}^N(X_i^L X_i^R + Y_i^L Y_i^R  - Z_i^L Z_i^R ). 
\end{align}

Unlike in the SYK Lindbladian, the size operator for the spin-chain is not proportional to $H_I= -2\mu (\mathcal{S}_Y + \mathcal{S}_Z)$, so its calculation is of more interest. 
Indeed, the
results depicted in the top panels of Fig.~\ref{fig:spin_size} show the existence of the $U(1)$ scars and that their size is markedly different from the non-scar states and also different from those in the SYK model.
The variance of the operator size vanishes only for the $U(1)$ scars, so it  can be employed to identify scars. As was expected, we do not observe $H^L$ scars. Moreover, we do not observe the degenerate scars reported in the SYK model (green circles).

\begin{figure*}[t!]
	\includegraphics[width=0.49\textwidth]{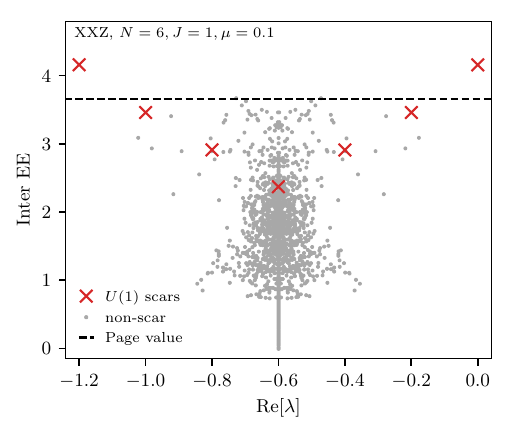}
	\includegraphics[width=0.49\textwidth]{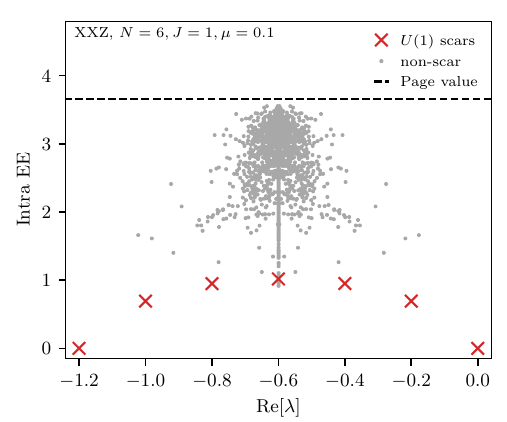}
	\caption{Intersite vs intrasite entanglement entropy Eq.~(\ref{eq:ee}) of eigenstates of the Liouvillian for the $N=6$ XXZ model Eq.~(\ref{eq:xxz_hamiltonian}). Results are qualitatively similar to those for the $U(1)$ scars of the SYK model. In both cases, the scar EE is in general different from the one for non-scar states, covers a relatively large range of values and, is sensitive to the partition.}
	\label{fig:spin_N=14_EE}
\end{figure*}

As in Sec.~\ref{sec:operator_size}, we define partial size operators similar to the even and odd size operators of  the Majorana fermion strings. Since $H_I = -2\mu (\mathcal{S}_Y + \mathcal{S}_Z)$, the expectation value of $\mathcal{S}_Y + \mathcal{S}_Z $
  is given by $-{\rm Re}[\lambda_k] /2\mu$ with $\lambda_k$ the eigenvalues of the vectorized Lindblad operator. All combinations of size operators  $\mathcal{S}_{\Sigma_i}$
  show a systematic behavior as a function of ${\rm Re}[\lambda_k]$ (which is termed secular variation in \cite{draayer1977strength,draayer1977strengthII}) as well as a fluctuating contribution. For $\mu =0.1$ we could not find a linear combination of size operators with a negligible smooth behavior. The combination with the smallest, but still significant, smooth behavior of the non-scar states is
 $ N/2 -\langle \mathcal{S}_X + \mathcal{S}_Z \rangle = \langle \sum_iY_i^LY_i^R\rangle/2$, see Fig.~\ref{fig:spin_size} (lower left).
    Note that
    $\langle \mathcal{S}_X + \mathcal{S}_Z \rangle
    = \langle \mathcal{S}_X - \mathcal{S}_Y \rangle  - \langle H_I \rangle/2\mu$, but
    $\langle \mathcal{S}_X - \mathcal{S}_Y \rangle$  has
    a stronger dependence on $\lambda$.  Because of the symmetry of Eq.~\eref{S2}, 
    $ N/2 -\langle \mathcal{S}_X + \mathcal{S}_Z \rangle$ 
is odd as a function of the real part of the eigenvalues of $\mathcal{L}+ N\mu/2$, and thus vanishes when the real part of
    these eigenvalues is equal to zero, which is the case for about $41\%$ of the states.
    The $U(1)$ scars     are a sum of strings of $\hat{Z}_i$, so naturally,
    $\langle {\mathcal S}_X\rangle$ and $\langle {\mathcal S}_Y\rangle$ are zero
    (and their difference is as well) so that $\langle \mathcal{S}_X + \mathcal{S}_Z \rangle
    = -{\rm Re}[\lambda_k]/2\mu$ for these states (red crosses in Fig. ~\ref{fig:spin_size}, bottom left).
    Therefore, this partial size operator is also suitable to characterize scars.

   The  bottom right panel of Fig.~\ref{fig:spin_size} shows
   a double-logarithmic plot of the full distribution of the normalized size discussed below
   in a small window around the center of the spectrum
   (main) and away from the center (inset).
As for the SYK model, see Fig.~\ref{fig:eohis},
we find a deviation from the ETH behavior, which predicts a Gaussian distribution.
However, according to the ETH, this is the case for an energy window with
a negligible smooth behavior  \cite{srednicki1994,draayer1977strength,draayer1977strengthII}. 
As was already discussed for the complex SYK model,
the local dependence of  the mean [$M(\lambda)$] and width [$\sigma(\lambda)$]
on the eigenvalue $\lambda$ has to be eliminated if there is a substantial variation over the energy window under consideration.
We have studied this dependence on both the real and imaginary parts of $\lambda$ by partitioning the support of the eigenvalues into 100 bins of equal
length, and calculating the mean and width for each bin. 
We did not find any significant smooth variation on the imaginary part of $\lambda$,
but there is an order of one smooth dependence on ${\rm Re}[\lambda]+\mu N$.
In the region $ |{\rm Re}[\lambda]+\mu N| < 0.2$, where a bin
contains at least 1000 eigenvalues (for 1000 realizations), the smooth dependence is
obtained by fitting  (piecewise) smooth functions,  $M(\lambda)$ and $ \sigma(\lambda)$,  to the values of the mean and
width for each bin. For example,   for $ |{\rm Re}[\lambda]+\mu N| < 0.02$,
we find
$M({\rm Re}[\lambda]) \approx 0.45 ({\rm Re}[\lambda]-\mu N)
+ (({\rm Re}[\lambda]-\mu N)/0.092)^3 $ and
$ \sigma({\rm Re}[\lambda]\approx 0.028 +5 |{\rm Re}[\lambda]-\mu N|$.  For
$|{\rm Re}[\lambda]+\mu N|>0.2$, both the width and the mean
show large bin-to-bin fluctuations and a reliable statistical analysis is not possible.
Because the operator $S_X +S_Z -N/2$ vanishes at ${\rm Re}[\lambda]=-\mu N/2$, to combine the data near this point into one histogram,
it is important to normalize the distribution of the sizes to
zero mean and unit variance.
This can be
achieved by considering the normalized expectation value,
\be
\langle S_X +S_Z\rangle_{\rm nor}=\frac{\langle S_X +S_Z\rangle -M(\rm{Re}[\lambda])
}{\sigma(\rm{Re}([\lambda])}.
\label{unf}
\ee
In Fig.~\ref{fig:spin_size} (bottom right), we show the distribution of the
normalized  sizes for $ |{\rm Re}[\lambda]+\mu N| < 0.0075$ (main) and
$0.057< |{\rm Re}[\lambda]+\mu N| < 0.075$
(inset).  Both have power-law tails, with a power that increases significantly away from the
center of the spectrum. For the complex SYK model we also found a power law
but a much weaker dependence on $\lambda$ (see Fig.~\ref{fig:eohis}).
The deviation from Gaussianity, which persists also after proper rescaling, Eq.~\eref{unf},  
suggests that ETH for Lindblad systems, if it applies, is quite different from the Hermitian analog.

Finally, Fig.~\ref{fig:spin_N=14_EE} shows the EE for the intersite and intrasite partitions.
  The results agree with the theoretical expectations and are qualitatively similar to
  those found in the Lindblad SYK model.

\section{Conclusion and outlook}
\label{sec:conclusion}

We have identified conditions for the existence of scarred eigenstates, which we termed Lindblad scars, of the vectorized Liouvillians that describe the dynamics of many-body quantum chaotic systems coupled to a Markovian bath. To illustrate the generality of our results, we have investigated in detail two models: a dissipative SYK model and a dephasing spin chain. In both cases, we constructed the scar states analytically. 

An explicit calculation of the operator size has revealed that many-body quantum scars
in these two models have an operator size given by the real part of the eigenvalues. Therefore, the size variance vanishes for scars and can be used to identify them. 
By contrast, non-scarred states show a smooth distribution centered around its average and with a power-like tail, which suggests strong deviations from the ETH in dissipative quantum chaotic systems. We have also observed that the entanglement properties of the many-body scars are
to some extent tunable and, for certain partitions,
the entanglement entropy of scarred states 
is close to the Page value, which suggests that these quantum scars are promising candidates to encode and transmit quantum information.

We have found that Lindblad scars can be determined from the parity or $U(1)$ symmetry of the Hamiltonian and the precise form of the jump operators. Therefore, a similar analysis for supersymmetric SYK models with one or more supercharges, together with a judicious choice of jump operators promises an even richer pattern of Lindblad scars. We have also identified additional scars in the complex SYK model that, to the best of our understanding, are not related to these symmetries.

It would be interesting to explore the precise role of these Lindblad scars in the quantum dynamics, see also Ref.~\cite{abanin2025}. We recall the many-body scars we have identified are not related to revivals because they are purely decaying modes, that is, they have purely real eigenvalues (see Ref.~\cite{jiang2025} for a recent account of scars with purely imaginary eigenvalues leading to revivals).
Another avenue for future research is a more detailed study of the size operator as a probe to characterize the ETH in dissipative many-body chaos for which the assumption of Gaussianity does not seem to apply.
   
\acknowledgments{Andrea Pizzi and Frank Schindler are thanked for useful discussions.
A. M. G. G. acknowledges support from the Natural Science Foundation of China (NSFC) through the Research Fund for International Senior Scientists No.\ 12350710180 and also the Individual Grant No.\ 12374138.
L.\ S.\ was supported by a Research Fellowship from the Royal Commission for the Exhibition of 1851.
J.\ J.\ M.\ V.\ is supported in part by U.S. DOE Grant No.\ DE-FG02-88ER40388.
A.\ M.\ G.\ G., L.\ S., and J.\ J.\ M.\ V.\ acknowledge hospitality and support from the Simons Center for Geometry and Physics and the program ``Non-Hermitian topology, geometry and symmetry across physical platforms'' where some of the ideas of this paper were developed.
}

\appendix
\section{Schmidt decomposition for the intrasite partition}
\label{appA}

In this appendix we show that also for the intrasite partition the EE can be expressed as in Eq.~\eref{entan}, and show that the intrasite EE
of the TFD state vanishes.

In this case, a state $  |k\rangle$ can be decomposed as (note the ordering of the indices)
\be
  |k\rangle = \sum_{i_1i_3; i_2i_4} K_{i_1i_2; i_3i_4}  |i_1i_3\rangle |i_2i_4\rangle,
  \ee
  where  $|i_1i_3\rangle\in \mathcal{H}^{L}_A\otimes\mathcal{H}^{L}_B$ and
$|i_2i_4\rangle\in  \mathcal{H}^{R}_A\otimes\mathcal{H}^{R}_B$.
  The singular value decomposition of $ K_{i_1i_3; i_2i_4}$ is given by
  \be
   K_{i_1i_2; i_3i_4} = \sum_{kl} \sigma_{kl} U_{i_1i_2; kl} V^\dagger_{kl; i_3i_4} 
   \ee
   with $ \sigma_{kl}$ the singular values, and $U$ and $V$ unitary matrices.  
     The matrix elements, $\langle j_1 j_2 |\rho_{\rm red}|j_1'j_2' \rangle$,
     of the reduced density matrix
$\rho_{\rm red}= \Tr_B |k\rangle\langle k| $
     can be expressed as
     \be
          && \sum_{j_3j_4}  \langle j_1j_3; j_2j_4 |k\rangle \langle k|j_1'j_3; j_2'j_4 \rangle\nn\\
     &=& \sum_{kl;k'l'}\sum_{j_3j_4}\sigma_{kl} U_{j_1j_2; kl} V^\dagger_{kl; j_3j_4}
\sigma_{k'l'} U_{j_1'j_2'; k'l'}^* V^T_{k'l'; j_3j_4}\nn\\        
     &=& \sum_{kl} \sigma_{kl}^2 U_{j_1j_3; kl}
 U_{j_1'j_3'; kl}^*,        
 \ee
 where in the last equality we used the unitarity of $V$.
 The reduced density matrix is then given by
\be
 \rho_{\rm red} &=&  \sum_{kl}\sum_{j_1j_2;j_1'j_2'} \sigma_{kl}^2 U_{j_1j_2; kl}
 U_{j_1'j_2'; kl}^*   |j_1j_2\rangle \langle j_1' j_2'| \nn \\     
 &=&  \sum_{kl} \sigma_{kl}^2 |kl\rangle \langle kl |
 \ee
 with
 \be
|kl\rangle = \sum_{j_1j_2} 
U_{kl;j_1j_2}^T   |j_1j_2\rangle.
\ee
This shows that also for the intrasite partition the reduced density matrix is given by
Eq.~\eref{entan}.

Finally, let us use this to calculate the EE of the state $|0\rangle$.  Since this state correspond to the infinite temperature TFD, it has maximal entanglement between the $L$ and $R$ spaces; correspondingly, the intersite EE reaches the maximum value $\log D=\frac N2\log2$.
  If the TFD state,
\be
|0\rangle=  \sum_{i_1i_2;i_1i_2}  |i_1i_2\rangle |i_1i_2\rangle/\sqrt D \in
\mathcal{H}^{L}\otimes\mathcal{H}^{R},
\ee
is expanded according to  the intrasite  partition \eref{intrasite},
\be
|0\rangle=\sum_{i_1i_2; i_3i_4}  K_{i_1i_2;i_3i_4}^{\rm intra}|i_1i_3\rangle|i_2i_4\rangle,
\ee
only the matrix elements  $K_{i_1i_1;i_2i_2}^{\rm intra}=1/\sqrt D$ are nonzero.
This is a projection matrix with only one nonzero singular value resulting in a vanishing
intrasite EE as found numerically in  Fig.~\ref{fig:spin_N=14_EE}.
  Note that the argument must be modified
for parity reflected TFD state,  $P^L|0\rangle$, which must be perpendicular to the TFD state.

\bibliography{26_02_03_mbscars-ls.bib}

\end{document}